\documentclass[epsf,useAMS,usenatbib]{mn2e}
\usepackage[usenames,dvipsnames]{color}
\usepackage{float}  
\usepackage[pdftex]{graphicx}
\usepackage{xspace}
\usepackage{amssymb}

\usepackage{textcomp}

\usepackage{listings}

\newcommand{\RNum}[1]{\uppercase\expandafter{\romannumeral #1\relax}}

\usepackage[T1]{fontenc}

\usepackage{subcaption}

\usepackage{titlesec}

\def\apjl{}
\def\pasa{Publications of the Astronomical Society of Australia}

\def\mnras{MNRAS}

\def\apj{ApJ}
\def\aap{A\&A}

\def\nat{Nature}

\newcommand{\apjs}{ApJS}		
\newcommand{\ssr}{Space~Sci.~Rev.}	

\newcommand{\myemail}{repetto@physics.technion.ac.il}
\newcommand{\beq}{\begin{equation}}
\newcommand{\eeq}{\end{equation}}
\newcommand{\mr}{\mathrm}
\usepackage{amssymb,amsmath}
\usepackage{epsfig}
\usepackage{rotating}
\usepackage{times}
\usepackage{natbib,twoopt}

\usepackage[breaklinks=true, colorlinks=true]{hyperref} 

\hypersetup{colorlinks=True, linkcolor=Magenta, citecolor=Orange}

\title[The Galactic distribution of X-ray binaries]
  {The Galactic distribution of X-ray binaries and its implications for compact object formation and natal kicks}
\author[S.~Repetto et al.]
  {Serena~Repetto$^{1,2}$\thanks{E-mail:
\myemail},~~Andrei~P.~Igoshev$^1$,  ~~Gijs~Nelemans$^{1,3}$\\
    $^1$Department of Astrophysics/IMAPP, Radboud University, P.O. Box 9010, 6500 GL Nijmegen, The Netherlands\\
    $^2$Physics Department, Technion - Israel Institute of Technology, Haifa, Israel 32000\\
    $^3$ Institute for Astronomy, KU Leuven, Celestijnenlaan 200D, 3001 Leuven, Belgium
  }

\date{\today}

\pagerange{\pageref{firstpage}--\pageref{}} \pubyear{0000}
\def\LaTeX{L\kern-.36em\raise.3ex\hbox{a}\kern-.15em
    T\kern-.1667em\lower.7ex\hbox{E}\kern-.125emX}

\begin{document}
\label{firstpage}

\maketitle

\begin{abstract}
The aim of this work is to study the imprints that different models for black hole (BH) and neutron star (NS) formation 
have on the Galactic distribution of X-ray binaries (XRBs) which contain these objects.
We find that the root mean square of the height above the Galactic plane of BH- and NS-XRBs is a powerful 
proxy to discriminate among different formation scenarios,
and that binary evolution following the BH/NS formation does not significantly affect the Galactic distributions of the binaries.
We find that a population model
in which at least some BHs receive a (relatively) high natal kick fits the observed BH-XRBs best.
For the NS case, 
we find that a high NK distribution,
consistent with the one derived from the measurement of pulsar proper motion,
is the most preferable.
We also analyse the simple method we previously used to estimate the minimal peculiar velocity of an individual BH-XRB at birth.
We find that this method may be less reliable in the bulge of the Galaxy for certain models of the Galactic potential, 
but that our estimate is excellent for most of the BH-XRBs.
\end{abstract}


\section{Introduction}
The formation mechanism of compact objects, neutron stars (NSs) and black holes (BHs),
is an unsolved problem in high-energy astrophysics.
A model for the formation of such objects
requires to perform physically-motivated simulations of the core-collapse supernova,
which is computationally challenging (see e.g. \citealt{2002ApJ...574L..65F}; \citealt{2012ApJ...759....5B}; \citealt{2012ARNPS..62..407J}).
Another possible way to investigate the formation of NSs and BHs
is to study the birth and evolution of X-ray binaries (XRBs) hosting a BH or a NS accreting from a stellar companion.
The orbital parameters, peculiar velocities
and Galactic position
of these binaries
directly follow from their evolutionary history,
and are affected in particular by the conditions at the moment of compact object formation
(see e.g. \citealt{1995MNRAS.274..461B}; \citealt{1998ApJ...504..967K}; \citealt{1999A&A...352L..87N}; \citealt{2007ASPC..367..533N}).

The measurement of pulsar proper motions (see e.g. \citealt{1994Natur.369..127L}; \citealt{1997MNRAS.291..569H}; \citealt{1997A&A...322..127H}; \citealt{2005MNRAS.360..974H}),
combined with the study of NS-XRBs
(e.g. \citealt{1992ApJ...387L..37J}; \citealt{1994ApJ...423L..43K}; \citealt{1997ApJ...489..244F}; \citealt{2000MNRAS.317..438K}; \citealt{2002ApJ...574..364P}),
has exposed evidence that some NSs receive a low velocity,
whereas others a high velocity
at formation (so called {\emph{natal kicks}}, NKs).
{{The prevailing idea is that NSs are formed either in a standard core-collapse
supernova (SN), or in a less energetic type of SN expected for star with small cores.
The latter can take place either as
an electron-capture SN or as an iron core-collapse SN with a small
iron-core mass  (\citealt{2004ApJ...612.1044P}; \citealt{2013ApJ...771...28T}; \citealt{2015MNRAS.451.2123T}; \citealt{2016arXiv161107562J}).}}
For the case of BHs, observations are rather scarce and patchy,
thus it is not yet possible to discriminate between different models of BH formation (\citealt{2003Sci...300.1119M}; 
\citealt{2004MNRAS.354..355J}; \citealt{2005ApJ...625..324W}; \citealt{2007ApJ...668..430D}; \citealt{2009ApJ...697.1057F}; \citealt{2009MNRAS.394.1440M}; \citealt{2012ApJ...747..111W}; \citealt{2014ApJ...790..119W}; \citealt{2015MNRAS.453.3341R}; \citealt{2016MNRAS.456..578M}).
In this paper, one of our goals is to investigate whether the observed Galactic distribution of XRBs hosting a BH (BH-XRBs) can
reveal something about how BHs are formed. The main underlying idea is that any
offset of a BH-XRB from the Galactic plane (assumed as birth place) is a signature of some peculiar velocity of the system with respect to the circular Galactic motion. 
The magnitude of such velocity gives clues on the SN mechanism,
 in particular on the magnitude of the NK at birth (\citealt{2004MNRAS.354..355J}; \citealt{2012MNRAS.425.2799R}). {{The idea of using the Galactic position and/or line of sight velocities of a population of XRBs to investigate the formation of compact objects was employed previously for the NS case (see e.g. \citealt{1995MNRAS.274..461B}; \citealt{1996ASIC..477..385J}).}}\\
 \indent We covered the topic of BH formation in two previous works. In~\citet{2012MNRAS.425.2799R},
we followed the Galactic trajectories of a simulated population of BH-XRBs,
and investigated which NK distribution gives rise to the observed $z$-distribution of BH-XRBs (where $z$ is the height above the Galactic plane).
The aim was to discriminate between {\emph{high}} and {\emph{reduced}} NKs for BHs.
High NKs are larger than the NK expected in a standard formation scenario for BHs,
in which the BH forms via fallback of material on to the proto-NS and the NK is caused by asymmetries in the SN ejecta.
In the standard scenario, the NK would conserve the linear-momentum and roughly scale as the NK received by the NS multiplied by the ratio between the mass of the BH and the mass of the NS. We call these kicks
as {\emph{reduced}} or {\emph{momentum-conserving}} NKs.
If the NS receives a NK of the order of $300$ km/s,
a $10~M_\odot$ BH would get a NK of $\approx 40$ km/s. We define {\emph{high}} NKs
as $\gtrsim 100$ km/s.
In~\citet{2012MNRAS.425.2799R}, we found that high NKs, comparable to NS NKs, were required.
In~\citet{2015MNRAS.453.3341R}, we combined the information from
the kinematics and binary evolution 
of a subset of BH-XRBs
to find evidence both for low and high NKs. In this Paper,
we aim at complementing and extending those previous studies.
Following up on the work by \citet{1995ApJ...447L..33V} and \citet{1996ApJ...473L..25W},
\citet{2004MNRAS.354..355J}
found that the root-mean-square (rms) value of the height above the Galactic plane of BH-XRBs is similar to that of NS-XRBs,
suggesting that BHs could also receive a {\emph{high}} kick at formation, or even one as high as NSs. In this work,
we develop this idea further.
We build synthetic populations of BH- and NS-XRBs and we model their
binary evolution and their kinematics in the Galaxy, to 
investigate whether 
different assumptions on compact object formation
(such as a different distribution for the NK and/or a different amount of mass ejected in the SN)
have an imprint on the observed Galactic distribution of BH- and NS-XRBs,
and we quantify these effects.

\indent Furthermore, we will dedicate part of this work to discuss a method we previously employed 
to calculate the minimum peculiar velocity at birth of individual BH-XRBs (\citealt{2012MNRAS.425.2799R};
\citealt{2015MNRAS.453.3341R}). 
The difference of the Galactic potential value between the observed position $(R, z)$ and 
its projection on to the Galactic plane was used to analytically derive a lower limit for the peculiar velocity at birth.
This method has been recently challenged by \citet{2016MNRAS.456..578M}.
We investigate how robust our estimate is,
i.e. how close this estimate is to the true value of the minimal peculiar velocity at birth,
how this estimate scales with the distance from the Galactic centre,
and how it varies for different choices of the Galactic potential.\\

The paper is structured as follows. In Section \ref{sec:estimate} we study our estimate for the peculiar velocity at birth of individual BH-XRBs.
In Section \ref{sec:BPS} we build synthetic populations of BH- and NS-XRBs for different assumptions on the compact object formation.
In Section \ref{sec:results}
we look at the Galactic distributions of these synthetic binaries
while investigating how they differ,
and inferring which NK distribution fits best the observed Galactic position of NS- and BH-XRBs.
In Section \ref{sec:Discussion} 
we discuss our findings and in Section \ref{sec:conclusion}
we draw our conclusions.

\begin{figure}
\centering
\includegraphics[width=\columnwidth]{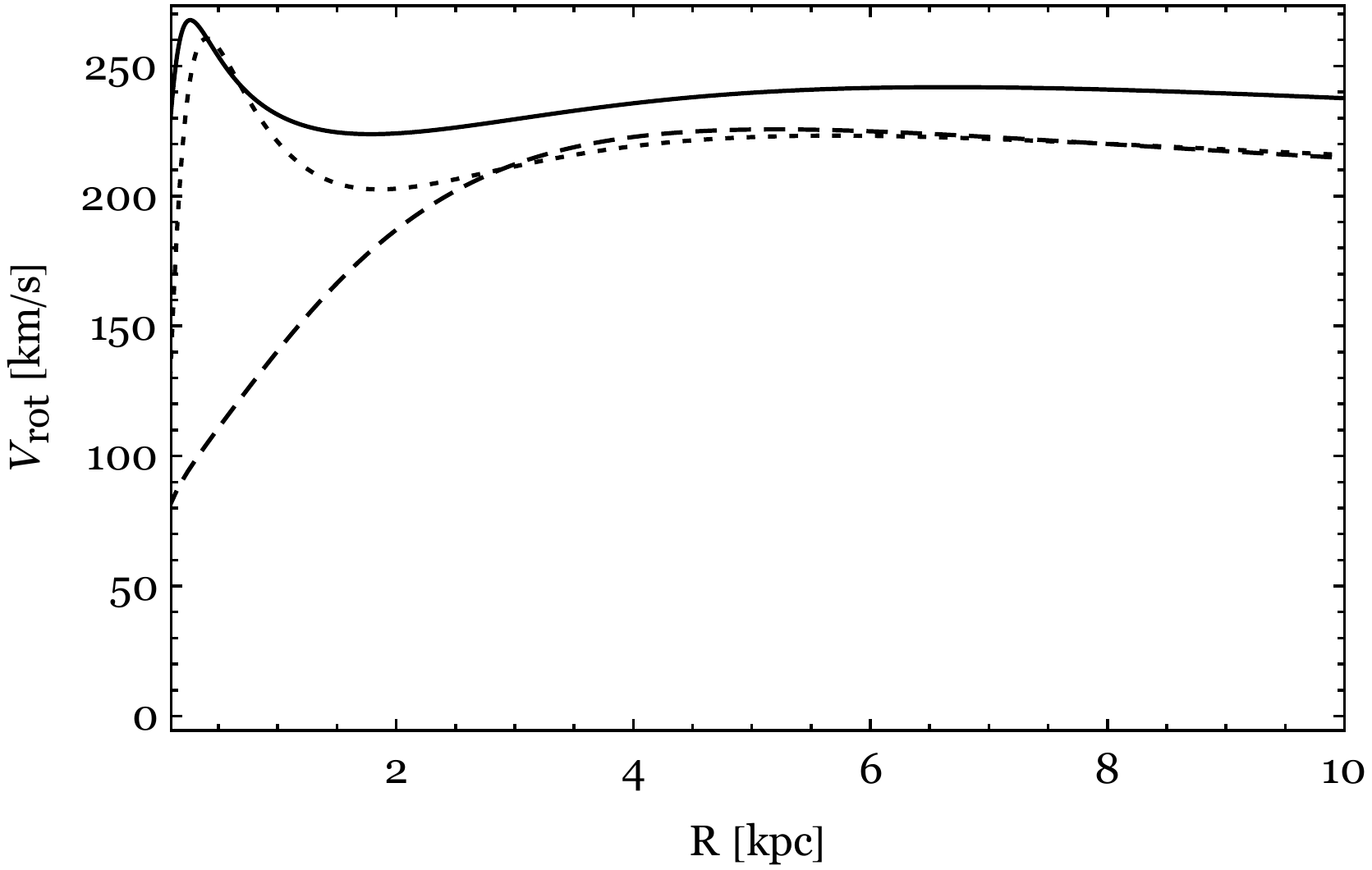}
\caption{Rotation curve for the Galactic potentials used in this work:
\citet{2015ApJS..216...29B} (dashed line), \citet{1990ApJ...348..485P} (dotted line), \citet{2013A&A...549A.137I} (solid line).}
\label{fig:RotCurve}
\end{figure}

\begin{figure*}
    \centering
    \begin{subfigure}[t]{0.33\textwidth}
        \centering
        \includegraphics[width=\columnwidth]{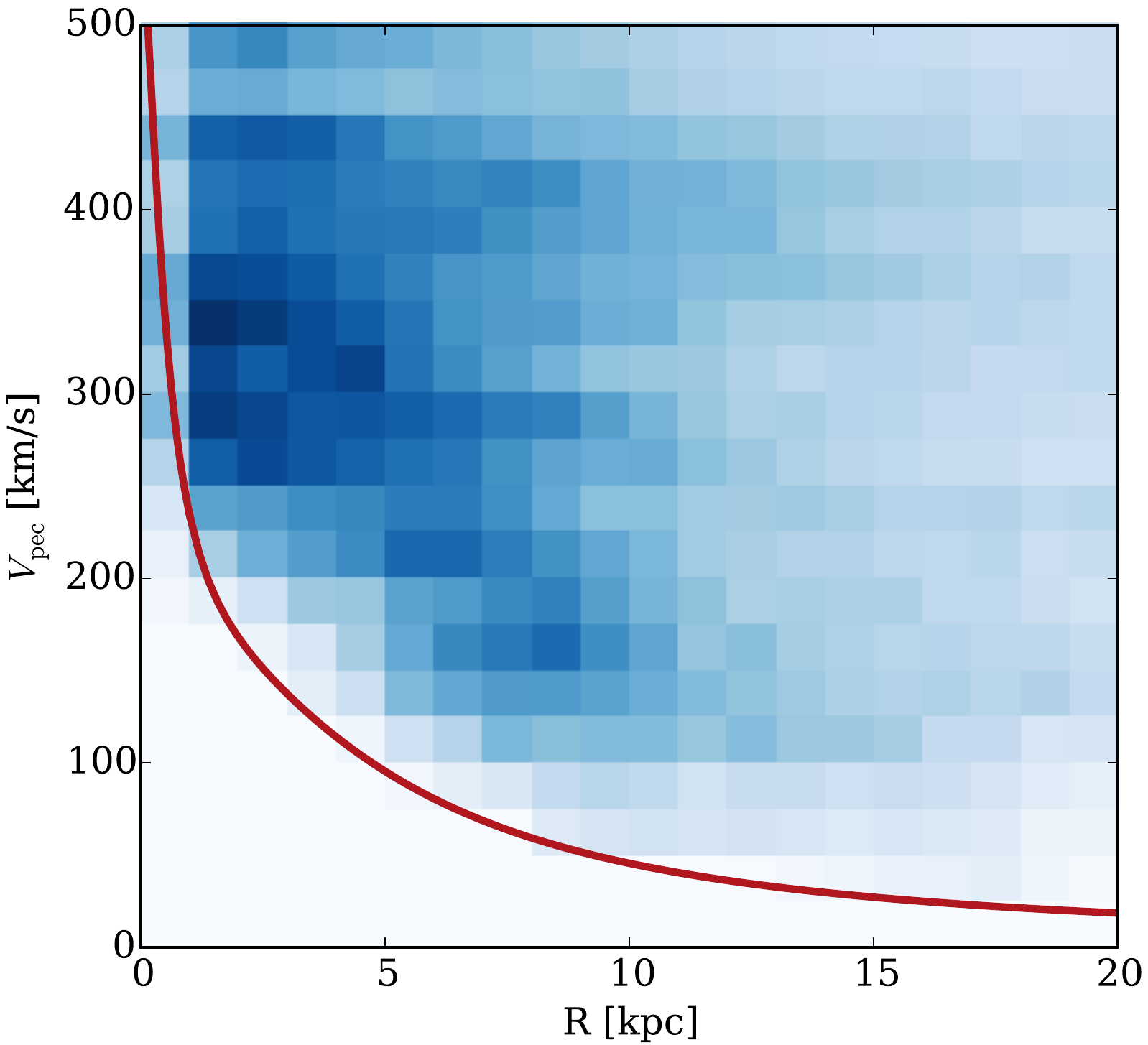}
    \end{subfigure}
    \begin{subfigure}[t]{0.33\textwidth}
        \centering
        \includegraphics[width=\columnwidth]{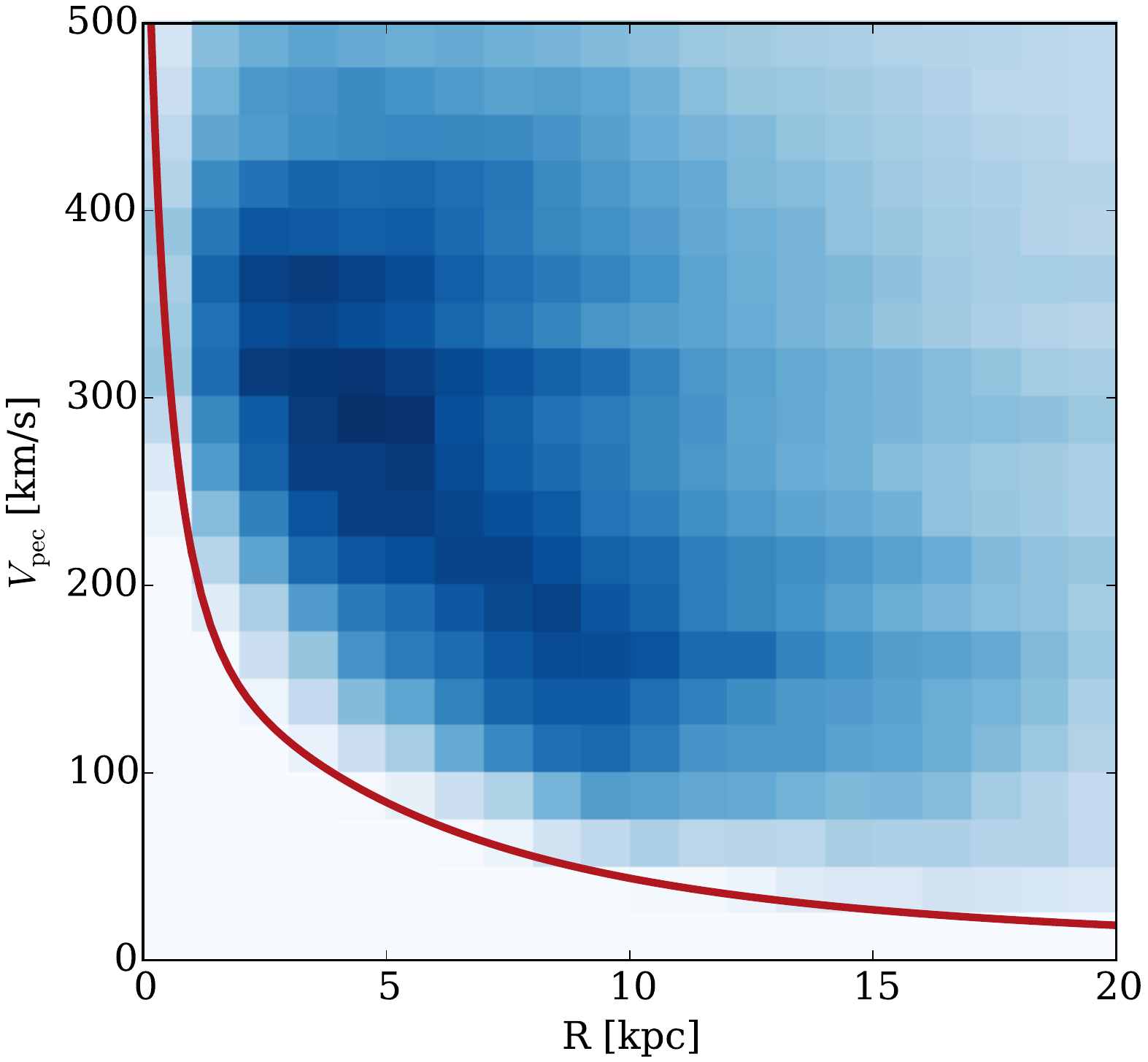}
    \end{subfigure}
        \begin{subfigure}[t]{0.33\textwidth}
        \centering
        \includegraphics[width=1.23\columnwidth]{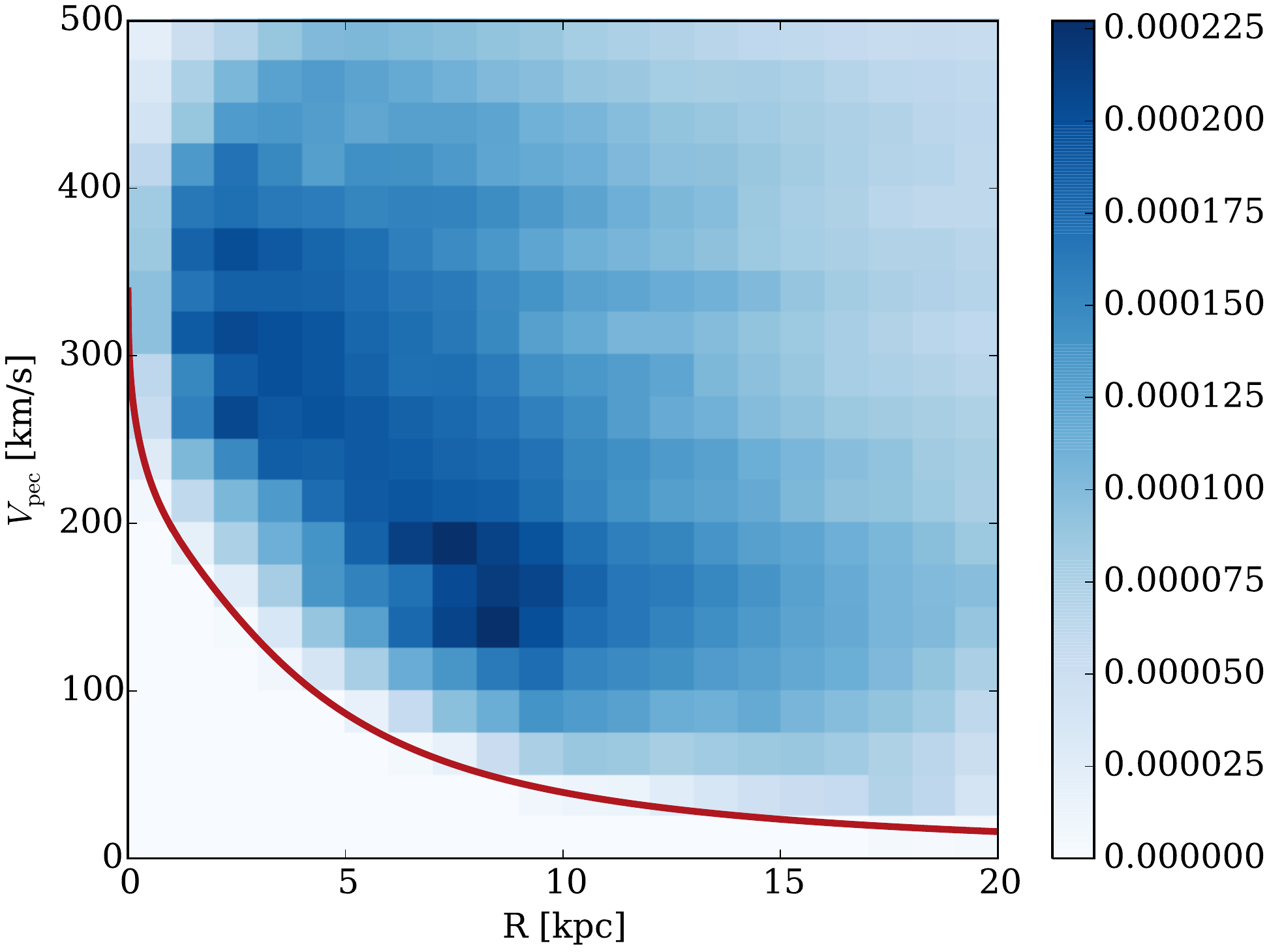}
    \end{subfigure}
    \caption{Density plots showing the fraction of systems in every bin of initial peculiar velocity $V_\mr{pec}$ and distance from the Galactic centre $R$ (projected on to the Galactic plane)
of points which reach a height above the Galactic plane greater than $1$ kpc. 
The red line shows our analytical estimate $V_\mr{pec, min}$.
We use three different potentials; from left to right: \citet{1990ApJ...348..485P}, \citet{2013A&A...549A.137I},~\citet{2015ApJS..216...29B}.}
\label{fig:density}
\end{figure*}

\section{Intermezzo}
\label{sec:estimate}
\subsection{On the estimate of the peculiar velocity at birth}
XRBs are thought to originate from binary progenitors born in the Galactic plane,
the birth-place of most massive stars
(\citealt{1995MNRAS.274..461B}).
When the compact object forms,
the binary typically acquires a peculiar velocity.
The mass ejection in the SN imparts a recoil velocity to the binary;
the NK adds up vectorially to this velocity, giving
the total peculiar velocity of the binary, $\vec{V}_\mr{pec}$.
Such a systemic velocity adds up vectorially to the local Galactic rotation and probably has no preferential orientation.
The full 3D velocity is measured only for a handful of BH-XRBs (see \citealt{2014PASA...31...16M}).
For these, the integration of the orbit backwards in time can in principle provide an estimate
for $\vec{V}_\mr{pec}$ at birth. However, uncertainty in the distance and differences in the Galactic potential
can prevent a unique determination of the initial position (see e.g. \citealt{2009ApJ...697.1057F}; \citealt{2009MNRAS.394.1440M}).
When the full 3D peculiar velocity is not known,
one can estimate $\vec{V}_\mr{pec}$ at birth using a simple model.
For an object located at Galactic height $z$\footnote{Throughout this work, we use a reference frame centered at the Galactic centre and cylindrical coordinates with $R$: the distance from the Galactic centre, and $z$: the height above the Galactic plane.}, we expect a trajectory purely perpendicular to the plane to be the one which minimises
the initial $\vec{V}_\mr{pec}$.
In our previous works~\citet{2012MNRAS.425.2799R} and~\citet{2015MNRAS.453.3341R},
we estimated the minimum peculiar velocity at birth of a BH-XRB
employing energy conservation along such trajectory,
and assuming that the maximum height $z$ from the plane
is the observed one.
We get:
\begin{equation}
V_{\mr{pec, min}}=\sqrt{2[\Phi\left (R_0, z \right ) - \Phi\left (R_0, 0
    \right )]}, 
    \label{eq:estimate}
    \end{equation}
where $\Phi \left (R, z \right )$ is a model for
the Galactic potential,
$R_0$ is the measured distance of the binary
from the Galactic centre projected on to the Galactic plane,
and $z$ is the current height above the plane.

Recently \citet{2016MNRAS.456..578M} argued that the difference in the gravitational potential between the observed
location and its projection on to the Galactic plane is not
an accurate estimate of the required minimum peculiar velocity at birth.
He suggests that there are always possible trajectories different from a purely perpendicular one
which require a lower ${V}_\mr{pec}$ at birth than the one estimated through equation \ref{eq:estimate}
to reach the same offset from the Galactic plane.

We check the validity of our estimate for the peculiar velocity at birth, $V_{\mr{pec,min}}$, for high-$z$ sources, 
performing a Monte Carlo simulation using the Python package for galactic dynamics {\tt{\small galpy}}\footnote{Available at {\tt{https://github.com/jobovy/galpy}}} (\citealt{2015ApJS..216...29B}). 
We simulate $1.1\times10^7$ points,
whose initial conditions are set as follows:
1) the initial position is at $(R,z)=(R_i,0)$,
where $R_i$ is uniformly distributed between $0$ and $18$ kpc;
2) the initial peculiar velocity $V_\mr{pec}$ is uniform between $0$ and $500$ km/s;
3) the orientation of this velocity is uniformly distributed over a sphere. 
We note that since we are only interested in the minimum value of $V_\mr{pec}$,
the shape of the assumed velocity distribution is not important.
We add the circular motion in the Galactic disc to the 3D peculiar velocity $\vec{V}_\mr{pec}$.
We integrate the orbits in the Galaxy for $5$ Gyr,
using a 4th-order Runge-Kutta integrator,
and we check for conservation of energy over the trajectory
making sure that the relative error on the energy is less than $10^{-5}$ at the final step.
We record the positions $(R,z)$
$500$ times over the orbit sampling from constant time steps,
along with the initial peculiar velocity $V_\mr{pec}$.
From the simulated points, we select only those ones located at 
$z^2 > 1$ at the sampled times, to represent high-$z$ sources.
We perform the simulation for three different choices of the Galactic potential:
model 2 of \citet{2013A&A...549A.137I}\footnote{When referring to the \citet{2013A&A...549A.137I} Galactic potential, we will hereafter refer to their model 2.} , \citet{1990ApJ...348..485P}, and the {\tt\small{MWPotential2014}} potential from \citet{2015ApJS..216...29B},
which are all multi-component potentials consisting of disc, bulge, and halo.
The \citet{1990ApJ...348..485P} potential is made up of two Miyamoto-Nagai potentials for disc and bulge,
and one pseudo-isothermal potential for the halo. The \citet{2015ApJS..216...29B} potential is made up of a power-law density profile with an exponential cut-off for the bulge, a Miyamoto-Nagai Potential for the disc, and a Navarro-Frenk-White profile for the halo. The \citet{2013A&A...549A.137I} potential is composed of two Miyamoto-Nagai potentials and 
a Wilkinson-Evans potential for the halo.
We show the rotation curve of each of the three potentials in Figure \ref{fig:RotCurve}.
 \citet{2013A&A...549A.137I} is the potential used by \citet{2016MNRAS.456..578M};
\citet{1990ApJ...348..485P} is the one we adopted in 
\citet{2012MNRAS.425.2799R};
the {\tt\small{MWPotential2014}} is a realistic model for the Milky Way potential
favoured by \citet{2015ApJS..216...29B}. 
We present the results of this simulation in Figure \ref{fig:density}.
The red line is our estimate for the peculiar velocity taking $z=1$ kpc in equation \ref{eq:estimate} and it follows the lower edge of the simulated points.

Figure \ref{fig:density} shows
that our analytical estimate (eq. \ref{eq:estimate}) successfully describes
the value and trend of the minimal peculiar velocity as a function of the Galactocentric distance.

\begin{figure*}
\includegraphics[width=2\columnwidth]{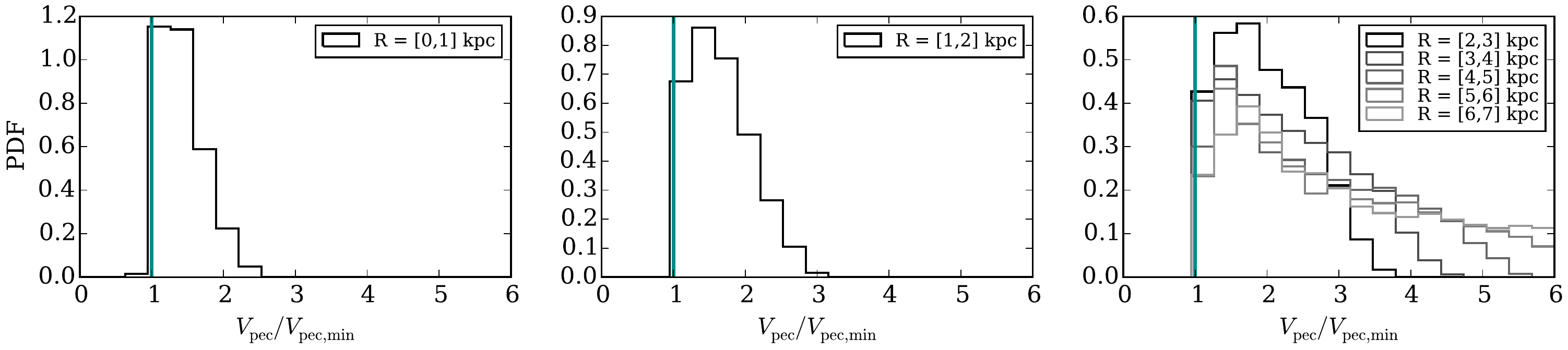}
\caption{Ratio $V_\mr{pec}/V_{\mr{pec,min}}$ for points such that the observed position is at $1<z<1.1$ kpc. Each panel shows a different R-bin. The Galactic potential is from \citet{2015ApJS..216...29B}.}
\label{fig:ratio1}
\includegraphics[width=2\columnwidth]{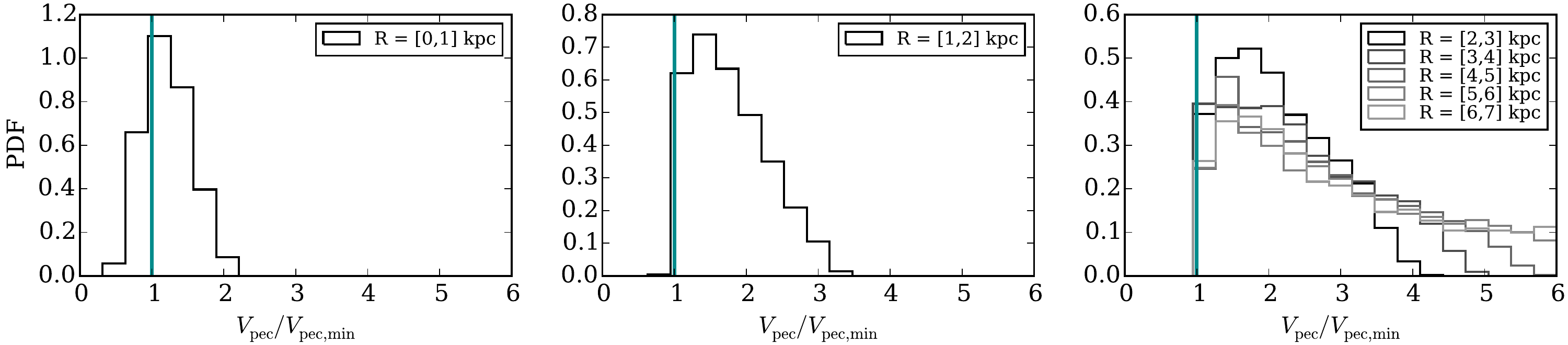}
\caption{Ratio $V_\mr{pec}/V_{\mr{pec,min}}$ for points such that the observed position is at $1<z<1.1$ kpc. Each panel shows a different R-bin. The Galactic potential is from \citet{2013A&A...549A.137I}.}
\label{fig:ratio2}
\includegraphics[width=2\columnwidth]{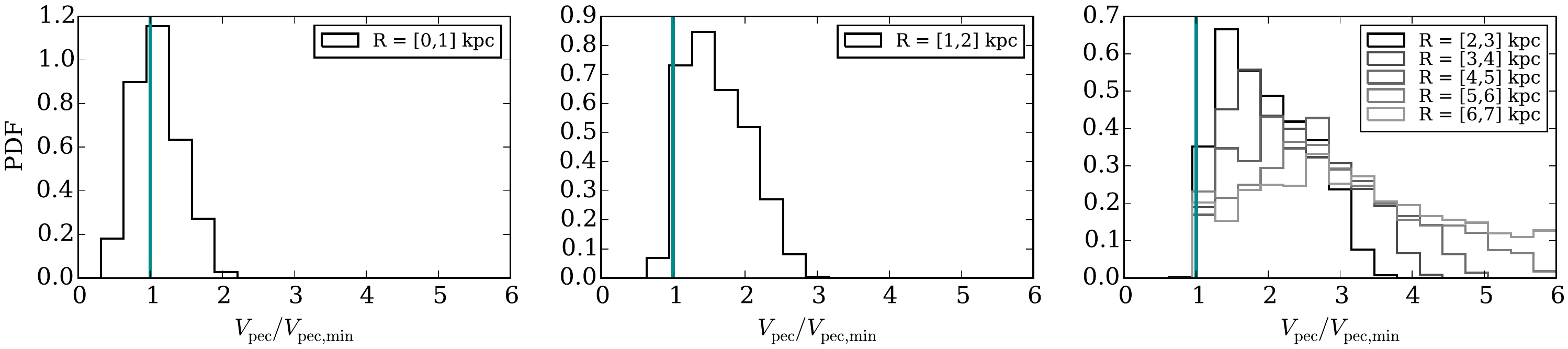}
\caption{Ratio $V_\mr{pec}/V_{\mr{pec,min}}$ for points such that the observed position is at $1<z<1.1$ kpc. Each panel shows a different R-bin. The Galactic potential is from \citet{1990ApJ...348..485P}.}
\label{fig:ratio3}
\end{figure*}

\begin{figure}
\centering
\includegraphics[width=0.9\columnwidth]{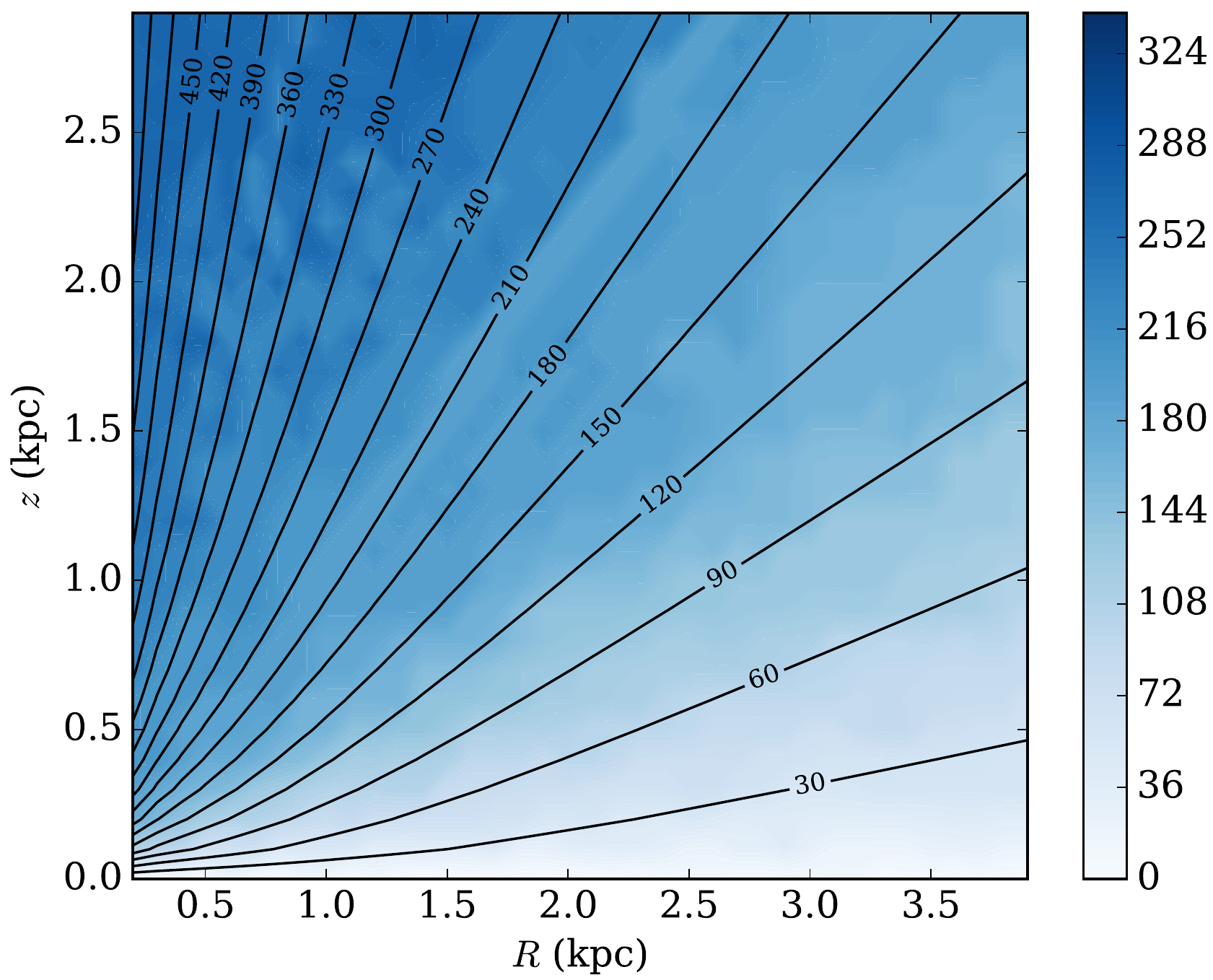}
\caption{Density map showing with color coding the minimal value for the peculiar velocity at birth $V_\mr{pec}$ of simulated points which reach that position $(R,z)$.
The contour lines show our analytical estimate $V_{\mr{pec,min}}$ at that position. The estimate differs strongly from 
the real value in the bulge region
(each solid line differs by $\pm 30$ km/s from the closest-neighbouring one).
The potential used is from \citet{2013A&A...549A.137I}.
}
\label{fig:Andrei}
\end{figure}

In order to better quantify the goodness of our estimator $V_{\mr{pec,min}}$, we compute the ratio $\gamma=V_\mr{pec}/V_{\mr{pec,min}}$ 
using $1$ kpc-wide bins in $R$, for those points which reach a
height above the Galactic plane along their orbit in the range $|z|=(1, 1.1)$ kpc.
The velocity $V_\mr{pec}$
is the actual initial peculiar velocity which we showed in Figure \ref{fig:density}.
We plot $\gamma$ in Figures \ref{fig:ratio1}, \ref{fig:ratio2}, \ref{fig:ratio3},
for the three different potentials.
$V_{\mr{pec,min}}$ is an excellent estimator for $R>1$ kpc,
since at these radii $\gamma$ is equal or greater than $1$.
It is less robust in the inner part of the bulge for the \citet{1990ApJ...348..485P} and \citet{2013A&A...549A.137I} potentials,
but not in the 
{\tt\small{MWPotential2014}} potential, that is fit to the most recent dynamical constraints on the Milky Way
and has a more realistic bulge model 
(Jo Bovy, private communication).
In the bulge region, our estimate is steeper than the real minimal peculiar velocity for the first two potentials,
i.e.,
it varies strongly for small variation in $R$.
This can be seen in Figure \ref{fig:Andrei},
where for every position $(R,z)$ we show as a density map the real minimal peculiar velocity at birth 
necessary to reach that position.
We integrated $10^4$ orbits for $5$ Gyr and using as potential the one in \citet{2013A&A...549A.137I}.
The contour lines show our analytical estimate $V_{\mr{pec,min}}$;
the discrepancy between the two velocities is evident in the inner region of the Galaxy.

Figures \ref{fig:ratio1}, \ref{fig:ratio2}, \ref{fig:ratio3} also show an increase of the average value of $\gamma$
with larger  distances $R$. This is an artefact caused by our choice of the $V_\mr{pec}$ initial distribution (uniform between $0-500$ km/s),
{{as the numerator in the ratio $\gamma$ can take all the values between $\approx V_{\mr{pec,min}}$ and $500$ km/s.}}

From our extensive analysis,
we find that the estimate $V_\mr{pec,min}$ accurately represents the real minimal value for the peculiar velocity at distances from the Galactic centre $\gtrsim 1$ kpc, and can be safely applied to estimate the peculiar velocity at birth of XRBs born in the Galactic plane.

\subsection{Effect of a different choice of the Galactic potential with an application to the observed BH-XRBs}
\label{sec:Effect}
\begin{figure}
\centering
\includegraphics[width=0.9\columnwidth]{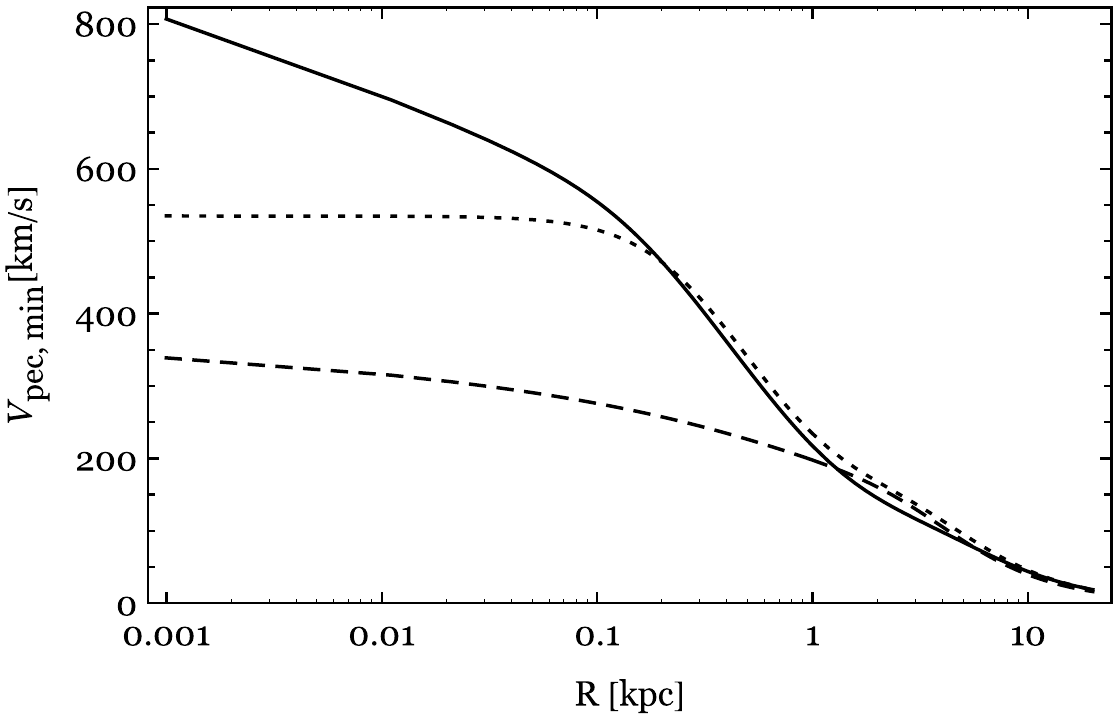}
\caption{Analytical estimate $V_{\mr{pec,min}}$ for the peculiar velocity at birth as a function of the distance from the Galactic centre $R$
(projected on to the Galactic plane) for the three different Galactic potentials used in this work:
\citet{2015ApJS..216...29B} (dashed line), \citet{1990ApJ...348..485P} (dotted line), \citet{2013A&A...549A.137I} (solid line).
We assumed $z=1$ kpc.
}
\label{fig:difference}
\end{figure}

The estimate $V_{\mr{pec,min}}$ is a function of the potential used,
in particular in the bulge, 
as can be seen in Figure \ref{fig:difference}, where we show $V_{\mr{pec,min}}$ for the \citet{1990ApJ...348..485P}, \citet{2013A&A...549A.137I} and 
\citet{2015ApJS..216...29B} potentials, and  assuming $z=1$ kpc in eq. \ref{eq:estimate}. Additionally,
from Figures  \ref{fig:ratio1}, \ref{fig:ratio2}, \ref{fig:ratio3},
we note that the fraction of systems with $\gamma<1$
in the region $R=[0,1]$ kpc
also strongly depends on the potential.
The minimum values $\gamma_\mr{min}$ are:
$1.01$, $ 0.72$, $0.61$ for \citet{2015ApJS..216...29B}, \citet{2013A&A...549A.137I}, \citet{1990ApJ...348..485P} potential respectively,
where these lower limits are defined such that $95\%$ of the points 
in the same bin have a value larger than the lower limit.

Figure \ref{fig:density} also shows that the Galactic bulge ($R\lesssim 1$ kpc)
is much less populated (an order of magnitude fewer systems than in regions at larger distance from the Galactic center).
There are two reasons for this: i) the bulge volume is small;
ii) it is unlikely for a binary born in the Galactic disc to overcome the strong potential well in its motion towards the Galactic bulge.
The inaccuracy of our analytical estimate in the bulge region affects only the source H1705-250,
which is the only BH-XRB located close enough to the Galactic centre (see Table \ref{tab:systems}), at $(R, z)\approx (0.5, 1.3)$ kpc (\citealt{1996ApJ...459..226R}).
Without a measurement of its 3D peculiar velocity, it is impossible to discriminate between a birth in the disc 
or a birth in the bulge (hence close to its observed position).
More in general, bulge sources are not suitable for estimating the peculiar velocities at birth,
since the current view on bulge formation is that it was not formed in situ.
The bulge population is thought to come from the disc through dynamical instabilities
(\citealt{2015ASPC..491..169G}),
with most of its mass coming from major and minor merger events with satellite galaxies
(\citealt{2011MNRAS.414.1439D}).\\
\indent We compute the minimum peculiar velocity at birth for the seven short-period BH-XRBs studied by \citet{2015MNRAS.453.3341R},
using the three Galactic potentials (see Table \ref{tab:minVpec}). We add to this sample
two other short-period BH-XRBs which we did not consider in \citet{2015MNRAS.453.3341R}
(XTE J1650-500 and XTE J1859+226),
due to the lack of a strong constraint on the BH mass (\citealt{2014SSRv..183..223C}).
For H 1705-250, we put in parenthesis the velocity
$V_{\mr{pec,min}}$ multiplied by the factor $\gamma$ found above.

\begin{table}
\caption{{Minimum peculiar velocity at birth for short-period BH-XRBs. The velocities are estimated using three different Galactic potentials and are given in km/s. The numbers in parenthesis for H 1705-250 correspond to correcting
the estimates for the inaccuracy of our analytical estimate in the bulge of the Galaxy (see Text).
}}
\label{tab:minVpec}
\begin{tabular}{l c c c c}
\hline
\multicolumn{5}{|c|}{$V_{\mr{pec,min}}$ [km/s]}\\
Source& Bovy & Pac.& Irrgang & Repetto et al.  \\
& & & & 2015 \\
\hline
XTE J1118+480 & 62 & 70 &  68 & 72 \\
GRO J0422+32 &  20 &  25  &  22 & 25\\
GRS 1009-45 &   34 &   40  & 37 &41\\
1A 0620-00 &  8&    10  & 8 & 10\\
GS 2000+251 &   12 &   15  &  12 & 15\\
Nova Mus 91 &  44 &   51  &  46 & 52 \\
H 1705-250 &   259 (262) &   363 (158)  &   350 (186) & 402\\
XTE J1650-500 & 17 & 21 & 16 & -\\
XTE J1859+226 & 61 & 68 & 68 & -\\
\hline
\end{tabular}
\newline
\end{table}

\newpage
We have found an error in the halo component of the \citet{1990ApJ...348..485P} potential that we used for the computation
of $V_{\mr{pec,min}}$ in~\citet{2015MNRAS.453.3341R}. This mostly affects the bulge source H 1705-250,
whereas the other six sources
are not greatly affected (compare third and last column in Table \ref{tab:minVpec}).

Accounting for the thickness of the Galactic disc instead of assuming a birth place at $z=0$ does
not  significantly affect the minimal peculiar velocity
(see \citealt{2016ApJ...819..108B}).

\citet{2016MNRAS.456..578M} used the source H 1705-250 to conclude that the difference
in the Galactic potential between the observed position and the projection of this position on to the Galactic plane
is not a conservative estimate of the minimal initial velocity of the binary. They show an example of a trajectory for 
H 1705-250 which starts from the Galactic plane and ends at the observed position
for an initial velocity of $\approx 230$ km/s,
lower than the value provided by eq. \ref{eq:estimate} (see Table \ref{tab:minVpec}).
We agree with his conclusion,
but only as far as sources close (or in) the bulge are concerned. On the contrary,
for sources located at $R\gtrsim 1$ kpc, our analytical estimate perfectly matches the real minimal velocity.
In \citet{2015MNRAS.453.3341R} we used the high minimal velocity at birth for XTE J1118+480 and 
H 1705-250 to claim that at least two out of the seven BH-XRBs we considered were consistent with a high (or relatively high) NK.
This holds true with our current revision of the minimal velocities at birth,
and we find another BH-XRB that is potentially consistent with a relatively high NK: XTE J1859+226.

The velocities we have been dealing so far with are {\emph{minimal}} velocities,
and do not necessarily correspond to {\emph{expected}} ({\emph{realistic}})
velocities. In what follows, we study the latter.

\section{A Binary Population Synthesis of BH- and NS-XRBs}
\label{sec:BPS}
In this part of the work, instead of dealing with the {{minimal}} peculiar velocities, 
we deal with the {{expected}} peculiar velocities.
We perform a binary population synthesis study of BH- and NS-XRBs,
starting just before the BH/NS formation,
varying the conditions at the formation of the compact object. The goal is to investigate the impact that different BH and NS formation assumptions
have on the Galactic distribution of XRBs containing a NS or a BH.
We assume that the binaries are formed in the Galactic thin disc,
where most of the massive stars reside (\citealt{2014MNRAS.437.1791U}). In this study, we do not account for the possibility
that a few systems could have been formed in the halo 
(i.e. in star clusters that have now been dissolved), and neither of the possibility that a few systems could have been ejected from globular clusters (GCs)
via N-body interactions.
{{GCs seem to be very efficient in producing NS low-mass X-ray binaries (NS-LMXBs), as $10\%$ of all NS-LMXBs are found in globular clusters, which contain only $\sim 0.1\%$ of all the stars in the Galaxy  (\citealt{2005ApJ...631..511I}).}}
 Such an investigation is, however, outside the scope of this paper.

We take different models for the formation of the compact object.
{{The NK is drawn either from a Maxwellian distribution 
peaked at $40$ km/s (with $\sigma \approx 28$ km/s) representing a low-NK,
or from a Maxwellian distribution peaked at $100$ km/s (with $\sigma \approx 71$ km/s) representing a high-NK.}}
We assume a certain amount of mass ejection in the SN, $M_\mr{ej}$.
BHs are thought to form either via prompt collapse of the progenitor star or via partial fallback of the SN ejecta onto the proto-NS
(see \citealt{2001ApJ...554..548F}).
In our models, the progenitors of BHs either do not eject any mass at collapse,
or they eject $4~M_\odot$. Stars with a ZAMS-mass larger than $\approx 25 M_\odot$ are thought to leave a BH behind
(see e.g. \citealt{2001ApJ...554..548F}; \citealt{2006csxs.book..623T}). For a progenitor of mass $25-60~M_\odot$,
the helium core mass (which collapses into a BH) is between $\approx 8-11~M_\odot$ (\citealt{2008ApJS..174..223B}),
which motivates our (conservative) choice for $M_\mr{ej}$. 
For the previous models, we assume a BH mass of $8~M_\odot$ (which is the typical 
mass for BHs in our Galaxy; \citealt{2010ApJ...725.1918O}).
We also picture a higher-mass helium star ($M_\mr{He}=15~M_\odot$)
which directly collapses into a BH with no mass ejection.
For NSs,
the ejected mass is calculated as: $M_\mr{ej}=M_\mr{He}-M_\mr{NS}$,
where $M_\mr{He}$ is the helium core mass ($M_\mr{He}=[2.8-8]~M_\odot$, see \citealt{2006csxs.book..623T}), and $M_\mr{NS}=1.4~M_\odot$. 
For the BH case the models are:
\begin{itemize}
\item Model 1: high NK, $M_\mr{He}=8~M_\odot$, $M_\mr{ej}=0$; 
\item Model 2: low NK, $M_\mr{He}=8~M_\odot$, $M_\mr{ej}=0$;
\item Model 3: high NK, $M_\mr{He}=8~M_\odot$, $M_\mr{ej}=4$;
\item Model 4: low NK, $M_\mr{He}=15~M_\odot$, $M_\mr{ej}=0$. 
\end{itemize}
For the NS case the models are:
\begin{itemize}
\item Model 5: high NK, $M_\mr{ej}$ uniform between $[1.4, 6.6]~M_\odot$;
\item Model 6: low NK, $M_\mr{ej}$ uniform between $[1.4, 6.6] ~M_\odot$.
\end{itemize}

For all the models, we simulate $3\times 10^7$ binaries composed of the helium star (which core-collapses)
and a companion star of $1~M_\odot$. The pre-SN orbital separation is uniformly drawn in the range $a_\mr{min}-50~R_\odot$ with zero initial eccentricity, where $a_\mr{min}$ is the minimal orbital separation such that either one of the two components fills its Roche lobe.
We calculate the effect of the compact object formation on the orbital properties and on the kinematics of the binary 
(for more details on the method, see \citealt{2015MNRAS.453.3341R}).
In particular, 
the effect of the mass ejection together with the NK impart a peculiar velocity to the binary:
\begin{equation}
V_{\rm pec} = \sqrt{\left(\frac{M_{\rm BH}}{M^\prime}\right)^2 V_{\rm NK}^2 + V_{\rm MLK}^2 -2\frac{M_{\rm BH}}{M^\prime}V_{\rm NK,x}V_{\rm MLK}},
\label{eq:PEC}
\end{equation}
where $M^\prime$ is the total mass of the binary after the SN,
$V_{\rm NK}$ is the magnitude of the NK,
$V_{\rm NK,x}$ its component along the orbital speed of the BH progenitor,
and $V_{\rm MLK}$ is the {\emph{mass-loss kick}}:
\begin{equation}
V_{\rm MLK} = \frac{M_\mr{ej}}{M^\prime} \frac{M_\star}{M} \sqrt{{\frac{GM}{a}}},
\label{eq:massloss}
\end{equation}
the recoil the binary gets because of the instantaneous mass ejection $M_\mr{ej}$ ($M$ is the initial mass of the binary;
$M_\star$ is the mass of the companion; $a$ is the initial orbital separation).
We follow the evolution of the binaries under the coupling between tides and magnetic braking using
the method developed in \citet{2014MNRAS.444..542R},
and select those systems that start mass transfer (MT), i.e. become X-ray sources, while the donor is on the main sequence.

We choose the radial distribution of the binaries
to follow the surface density of stars in the thin disc:
$\Sigma (R) \sim \Sigma_0 \exp({-R/R_{\rm d}})$, 
with $R_{\rm d} \sim 2.6$ kpc (\citealt{2011MNRAS.414.2446M}; \citealt{2012ApJ...753..148B}), and with a maximum distance from the Galactic Centre of $R_\mr{max} = 10$ kpc.
Concerning the height above the plane, we model it as an exponential
with scale height $h$ equal to the scale height of the thin disc ($h=0.167$ kpc; \citealt{2008gady.book.....B}).
This is a conservative choice for the scale height,
being the scale height of massive stars in the disc typically smaller
($h \sim 30$ pc; see Table 4 in \citealt{2014MNRAS.437.1791U}).
We assume that the stars follow the Galactic rotation,
with no additional component.
{{Various mechanisms can heat up the stars in the disc, increasing their dispersion velocity, such as encounters with spiral density waves, giant molecular clouds, and various other forms of stochastic heating (\citealt{1981gask.book.....M}; \citealt{2002ASPC..275..281S}; \citealt{2004A&A...423..517R}; \citealt{2016MNRAS.462.1697A}). \citealt{2004A&A...423..517R}, using a large sample of late-type dwarfs in the Milky Way disc, measured a dispersion in the three velocity-components 
of $\sigma_u\approx 50$ km/s, $\sigma_v \approx 30$ km/s, $\sigma_w \approx 20$ km/s at $t \approx 5\times10^9$ Gyr (see also \citealt{2009A&A...501..941H}). We neglect this influence, as we expect that for low-mass stars hosted in (massive) binaries these velocity would be significantly lower.}}

We integrate the orbit of the binaries for $5$ Gyr
using the {\tt\small{MWPotential2014}} potential from \citet{2015ApJS..216...29B},
which is a realistic model for the Milky Way potential.
We record the position along the orbit every $5$ Myr after $1$ Gyr.

\subsection{Observational samples}
\subsubsection{Black Hole X-ray Binaries}
\label{sec:ObsSample}
Using the catalogue of \citet{2016A&A...587A..61C}, we classify the systems into three main groups:
\begin{enumerate}
\item short-period, dynamically confirmed BH-XRBs (9 systems);
\item short-period, dynamically confirmed BH-XRBs + short-period BH candidates (15 systems);
\item short- and long-period, dynamically confirmed BH-XRBs (12 systems),
\end{enumerate}
which we list in Table \ref{tab:systems}, along with their Galactic position $(R, z)$ derived from
their sky-position and distance.
Dynamically-confirmed BHs are those 
for which a dynamical measurement of the BH mass is available (see e.g. \citealt{2014SSRv..183..223C}).
\begin{table}
\caption{Galactic position of the three classes of BH-XRBs;
$R$ is the distance from the Galactic centre, $|z|$ is the absolute value of the height above the plane. In parenthesis we put the uncertainty on the measurements. See \citet{2016A&A...587A..61C} for the references for the distance measurements.}
\label{tab:systems}
    \begin{tabular}{*{3}{|l}|} 
        \hline Name & R & $|z|$  \\
        & \small{(kpc)} & \small{(kpc)} \\ \hline
        \multicolumn{3}{|c|}{short-period confirmed }  \\ 
       XTE J1118+480 & $ 8.74$ $(0.1)$ & $ 1.52$ $ (0.2)$ \\
GRO J0422+32    & $10.38  $ $(0.65)$ & $0.51  $ $(1.15)$\\
GRS 1009-45  & $8.49 $ $ (0.25)$  & $  0.62 $ $ (0.1) $ \\
1A 0620-00  & $8.93 $ $  (0.08) $& $0.12 $ $ (0.01) $ \\    
GS 2000+251 & $7.21 $ $ (0.3)$	& $  0.14 $ $ (0.08)$  \\
Nova Mus 91  & $ 7.63 $ $ (0.2)$ 	& $ 0.72 $ $ (0.1) $ \\
H 1705-250 & $0.53  $ $ (2.9) $	& $1.35  $ $ ( 0.85) $  \\
XTE J1650-500  & $5.71 $ $ (1.35)$	& $0.15  $ $ (0.075) $ \\
XTE J1859+226  &  $10.03 $ $ (3.05)$ & $1.87 $ $ (0.65)$  \\
\hline
  \multicolumn{3}{|c|}{long-period confirmed }  \\ 
XTE J1550-564  & $4.96 $ $ (0.15)$ & $0.14 $ $ (0.05)$ \\
GRS 1915+105  & $6.62 $ $   (0.99)$ & $0.03  $ $  (0.008)$ \\ 
GS 2023+338 (V404 Cyg)  & $7.65 $ $ (0.001)$ &   $0.09 $ $  (0.005)$ \\  
\hline
 \multicolumn{3}{|c|}{short-period candidates}  \\ 
        MAXI J1836-194 &  $2.08 $ $ (1.15)$	& $0.65 $ $  (0.25)$  \\
MAXI J1659-152   & $0.82 $ $ (1.55)$	& $2.45  $ $ (1.05)$  \\ 
XTE J1752-223  & $2.15 $ $  (1.55)$	& $0.22 $ $  (0.1)$  \\
SWIFT J1753.5-0127 & $3.64 $ $ (0.65)$	& $ 1.27 $ $ (0.45)$  \\
4U 1755-338 & $1.56 $ $ (1.8)$	& $0.55 $ $ (0.25)$  \\
GRS 1716-249 & $5.62 $ $ (0.4)$ & $0.29 $ $ (0.05)$  \\
\hline
    \end{tabular}
    \end{table}

The observed BH-XRBs are both long ($P_\mr{orb} > 1$ day) and short orbital period ($P_\mr{orb}\lesssim 1$ day),
thereby originating from different evolutionary paths.
Hence, in order to compare the observed systems with the simulated binaries,
we need to produce two separate synthetic population of binaries,
one population with short-period and one population with long-period,
to which we compare the observed binaries according to their type.
For the short-period binaries, we follow the binary evolution of simulated binaries using the method we explained in Section \ref{sec:BPS}.
For the long period ones, 
which are driven by the nuclear evolution of the donor,
we model them assuming the post supernova orbital separation to be such that
$a_\mr{circ}=a_\mr{postSN}(1-e^2)\leq 20~R_\odot$,
where $a_\mr{circ}$ is the circularised orbital separation and $e$ is the eccentricity in the post-SN configuration.
This assumption is based on the fact that long-period binaries evolve to longer and longer period during the MT phase,
hence: $a_\mr{circ}\approx a_\mr{MT,0}<a_\mr{MT, obs}$,
where $a_\mr{MT,0}$ is the orbital separation at the onset of MT,
and $a_\mr{MT, obs}$ is the observed orbital separation. The assumptions on the compact object formation are the same as for the short-period binaries,
as well as the masses of the binary components.
Since our simulated binaries have a companion mass of $1~M_\odot$, 
we exclude from the observed sample those binaries with a companion mass: $\gg 1~M_\odot$
(GRO J1655-40, 4U 1543-475, and SAX J1819.3-2525).

\begin{figure}
\centering
\includegraphics[width=0.9\columnwidth]{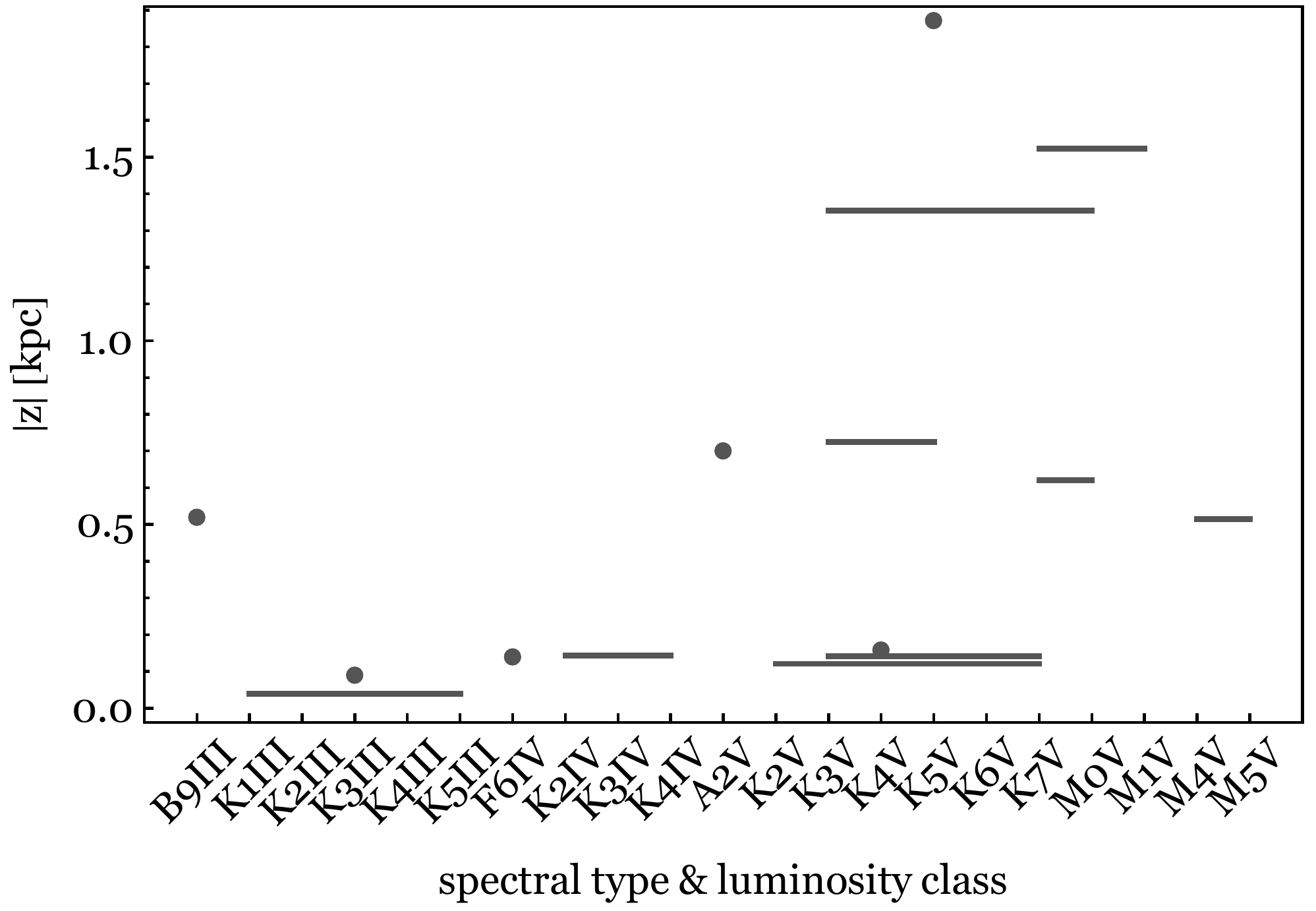}
\caption{The height above the Galactic plane $|z|$ and the spectral type and luminosity class of the 15 dynamically-confirmed BH-XRBs. When the spectral type of the donor star in the system is not univocally identified,
we indicate the range of possible types.}
\label{fig:SpecType}
\end{figure}

We account for a possible observational bias on the dynamically-confirmed BH-XRBs.
In order to get a dynamical measurement of the BH mass,
hence fully confirming the nature of the source,
high signal-to-noise optical spectra are required;
this might be prevented in regions of high extinctions,
i.e. in and close to the Galactic plane.
We then remove from our simulated populations
those binaries which are located at $z\leq 0.1$ kpc.
We note that the lowest $z$ in the sample of short-period dynamically confirmed BH binaries is 
for 1A 0620-00 ($z\approx -0.12$ kpc; see Table \ref{tab:systems}). 
For the long-period binaries,
we exclude from the study the sources GRS 1915+105 (donor spectral type: K1/5 III)
and V404 Cyg (donor spectral type: K0 IV), which are located at $z \approx -0.03$ kpc
and $z \approx -0.09$ kpc respectively (see Table \ref{tab:systems}).
These two systems do have a dynamical measurement of the BH mass (see \citealt{2014SSRv..183..223C}).
In Figure \ref{fig:SpecType}
we plot the absolute value of the height $z$ versus the spectral type and luminosity class of
the $15$\footnote{12 systems from Table \ref{tab:systems} to which we add the three BH-XRBs with an intermediate-mass companion.} dynamically-confirmed BH-XRBs (the spectral types are from \citealt{2016A&A...587A..61C}). At small $z$,
stars have an earlier spectral type and/or
are giants or sub-giants. Whereas MS/dwarf stars
tend to be seen at larger distances above the plane.

The only long-period binary in our sample,
after removing those sources close to the Galactic plane, is
XTE J1550-564, which has a current orbital separation
of $12~R_\odot$,
consistent with our assumption on $a_\mr{circ}$.

\subsubsection{Neutron Star X-ray Binaries}
The Galactic population of NS-XRBs consists of more than $30$ objects (see \citealt{2004MNRAS.354..355J}
and references therein). For our study, we select the $10$ ones with a short-orbital period
($P_\mr{orb}< 1 $ day; see Table 2 in \citealt{2004MNRAS.354..355J}).
The identification of a NS-XRB typically occurs via the detection
of X-ray bursts which ignite on the surface of the NS. Therefore, unlike for BHs,
there are potentially no biases against the identification of such systems.

\section{Results of the Binary Population Synthesis}
\label{sec:results}
\subsection{The expected vertical distribution of BH- and NS-XRBs}
\begin{figure}
\centering
\includegraphics[width=\columnwidth]{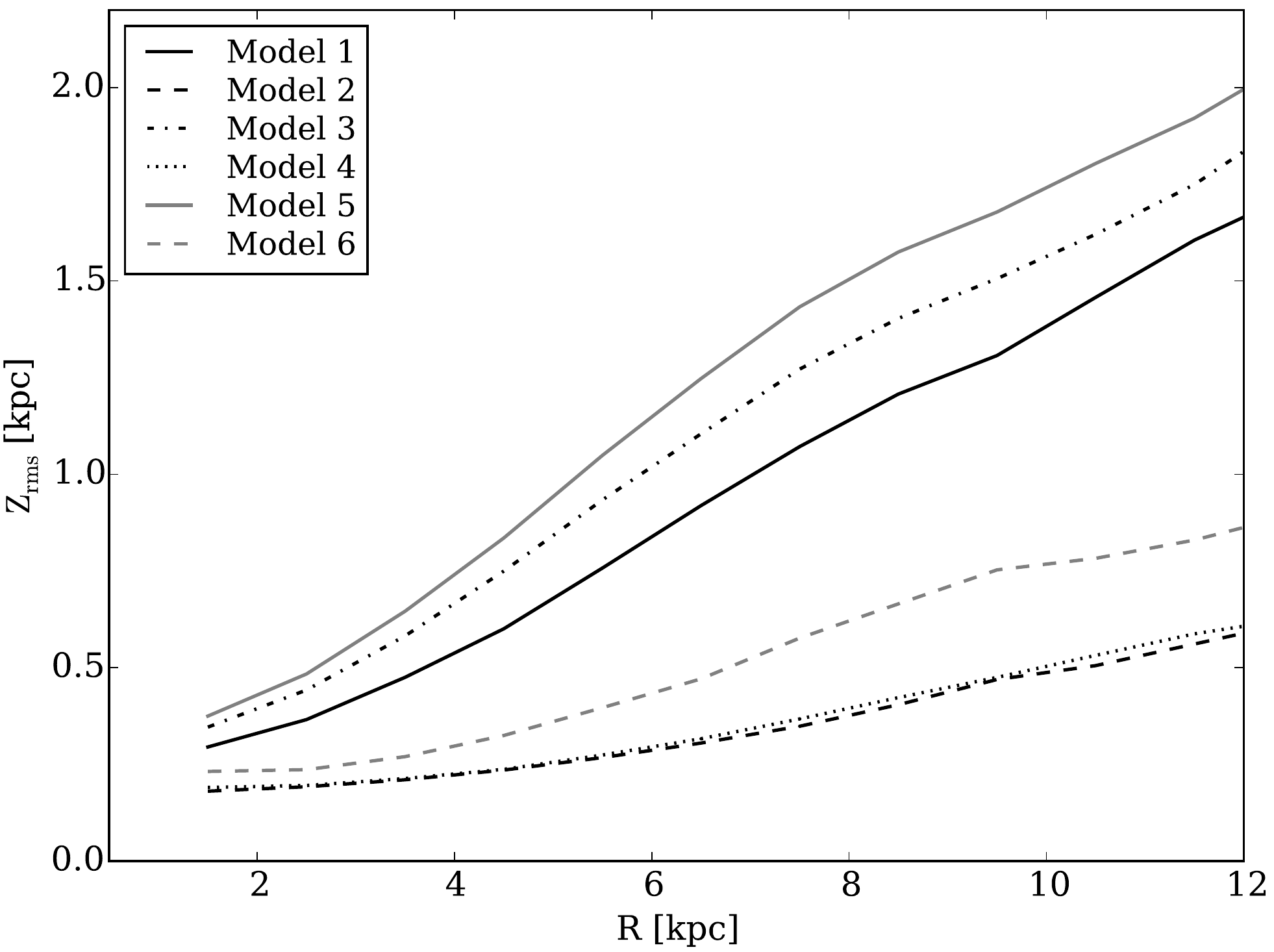}
\caption{Root-mean-square of the height $z$ above the Galactic plane of simulated BH- and NS-XRBs for the different models used.
}
\label{fig:zrms}
\end{figure}

\begin{figure}
\centering
\includegraphics[width=0.9\columnwidth]{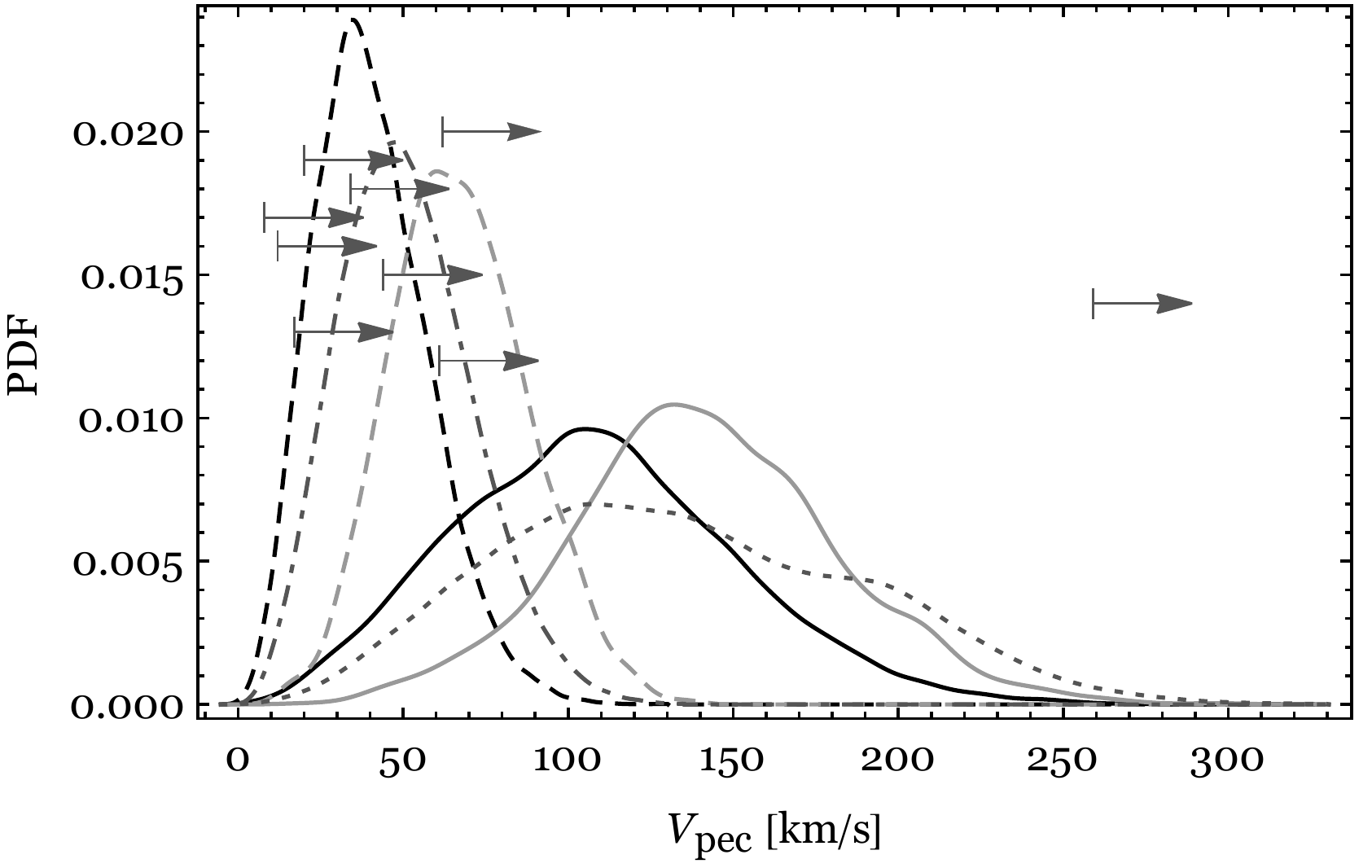}
\caption{Distribution of the peculiar velocity $V_\mr{pec}$ (after the formation 
of the compact object) of BH-XRBs in Model 1 (black solid line) and Model 2 (black dashed line),
and of NS-XRBs in Model 5 (grey solid line) and Model 6 (grey dashed line). The dotted and dotted-dashed dark-grey lines are variations of Model 1 (see Section \ref{sec:Discussion} for details).
The arrows represent the lower limits on the peculiar velocity at birth for the 9 short-period BH-XRBs using the potential
from \citet{2015ApJS..216...29B}.
}
\label{fig:vsys}
\end{figure}

The scale height of BH- and NS-XRBs is a proxy of the effect of different compact object formation mechanisms on to the Galactic distribution of the binaries.
We quantity the scale height of the binaries as the rms of their height $z$ as a function of $R$ for all points.
To plot the results, we bin the systems into $1$ kpc-wide bins in the $R$-direction. We show the results in Figure 
\ref{fig:zrms} for the six models. The monotonic rise of $z_\mr{rms}$ is expected,
since the Galactic potential becomes weaker further away from the Galactic centre,
and the binary moves further up for the same initial velocity.
It is interesting to note that
if BHs and NSs receive the same NK,
they would still show a different scale height,
with NSs reaching larger distances from the Galactic plane
(compare black solid line with grey solid line,
and black dashed line with grey dashed line).
This is due to the fact 
that for the same linear momentum, a binary with a larger mass receives a lower $V_\mr{pec}$ (as is shown in Figure \ref{fig:vsys}). 
If the progenitor of the BH ejects mass at core-collapse as in Model 3 (see black dashed-dotted line in Figure \ref{fig:zrms}),
it will move further out from the plane than when no mass is ejected,
since the mass ejection adds an extra contribution to $V_\mr{pec}$.
Furthermore, $V_\mr{pec}$ does not depend on the mass of the BH when no mass is ejected at BH formation
(black dashed and black dotted lines in Figure \ref{fig:zrms}),
since it scales as $V_\mr{pec}=\sqrt{\left (\frac{M_\mr{BH}}{M_\mr{BH}+M_2}\right )^2 V_\mr{NK}^2}\sim V_\mr{NK}$,
for low-mass companion stars (see equation \ref{eq:PEC}). 

In Figure \ref{fig:vsys} we also show as arrows the lower limits on the peculiar velocity at birth
of the 9 BH-XRBs we studied in Section \ref{sec:Effect}. It is clear that a high-NK distribution (darker-grey solid line) more easily accounts for the higher-velocity systems,
as $4$ systems lie in or
beyond the high-velocity tail of the distribution corresponding to the low-NK model.

\citet{2004MNRAS.354..355J} found a similar $z_\mr{rms}$ between NS- and BH-XRBs
and deduced that BHs should receive NKs too,
unless differences in the binary evolution and observational biases were strong.
We confirm that accounting for binary evolution does not strongly change the Galactic distributions of BH- and NS-XRBs.
However, the scale height does strongly depend on the position in the disc.

\begin{figure}
\centering
\includegraphics[width=0.9\columnwidth]{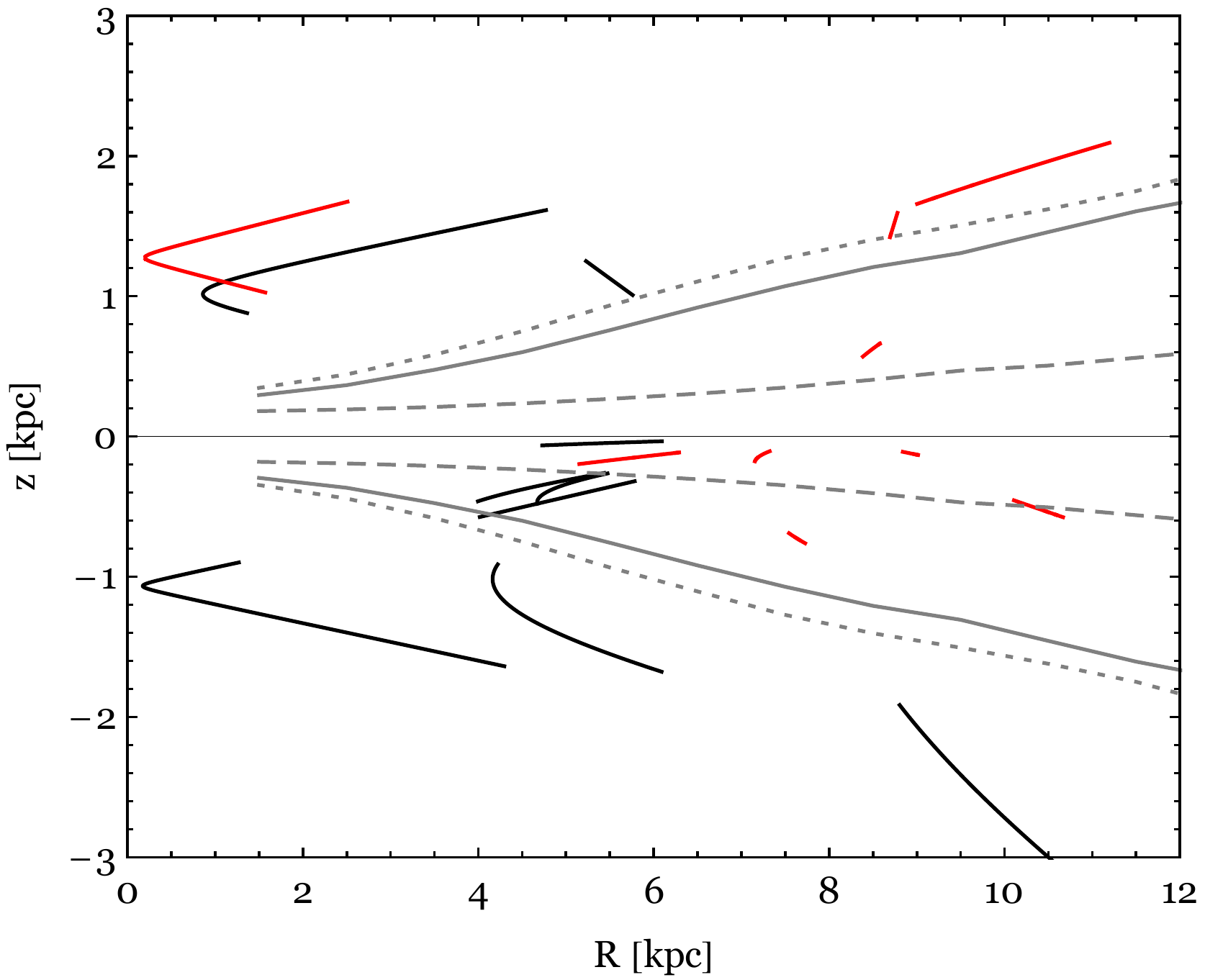}
\caption{Galactic distribution of BH-XRBs (red lines) and NS-XRBs (black lines). $R$ is the distance from the Galactic center projected on to the plane,
and $z$ is the height above the plane. One NS-XRB falls off the figure: XTE J2123-058. For each source, the line accounts for the uncertainty on the distance. We also show the results from the population study in terms of $z_\mr{rms}$ as a function of $R$: Model 1 (grey lines), Model 2 (grey-dashed lines), Model 3 (grey-dotted lines).}
\label{fig:Space}
\end{figure}

{{We compute the $z_\mr{rms}$ of the observed BH-XRBs,
both of the whole sample and of the dynamically-confirmed systems only.
We find a rms of $\approx 0.98 \pm 0.10$ kpc and $\approx 0.86 \pm 0.10$ kpc respectively. 
For the short-period NS systems, we calculate a $z_\mr{rms}$ of $1.24 \pm  0.06$ kpc,
when excluding the source XTE J2123-058 since its velocity is consistent with being a halo source,
as \citet{2004MNRAS.354..355J} noted. The error on these $z_\mr{rms}$-values accounts for the uncertainty on the distance to the sources.}}
In Figure \ref{fig:Space} we show the Galactic distribution of NS and BH systems (the lines account for the uncertainty in the distance to the source).
The result that NS systems should have a larger scale height than BH systems  is consistent with what the
observed populations show.

\subsection{The influence of the orbital separation distribution of the binary progenitors}
\begin{figure} 
\centering
\includegraphics[width=0.9\columnwidth]{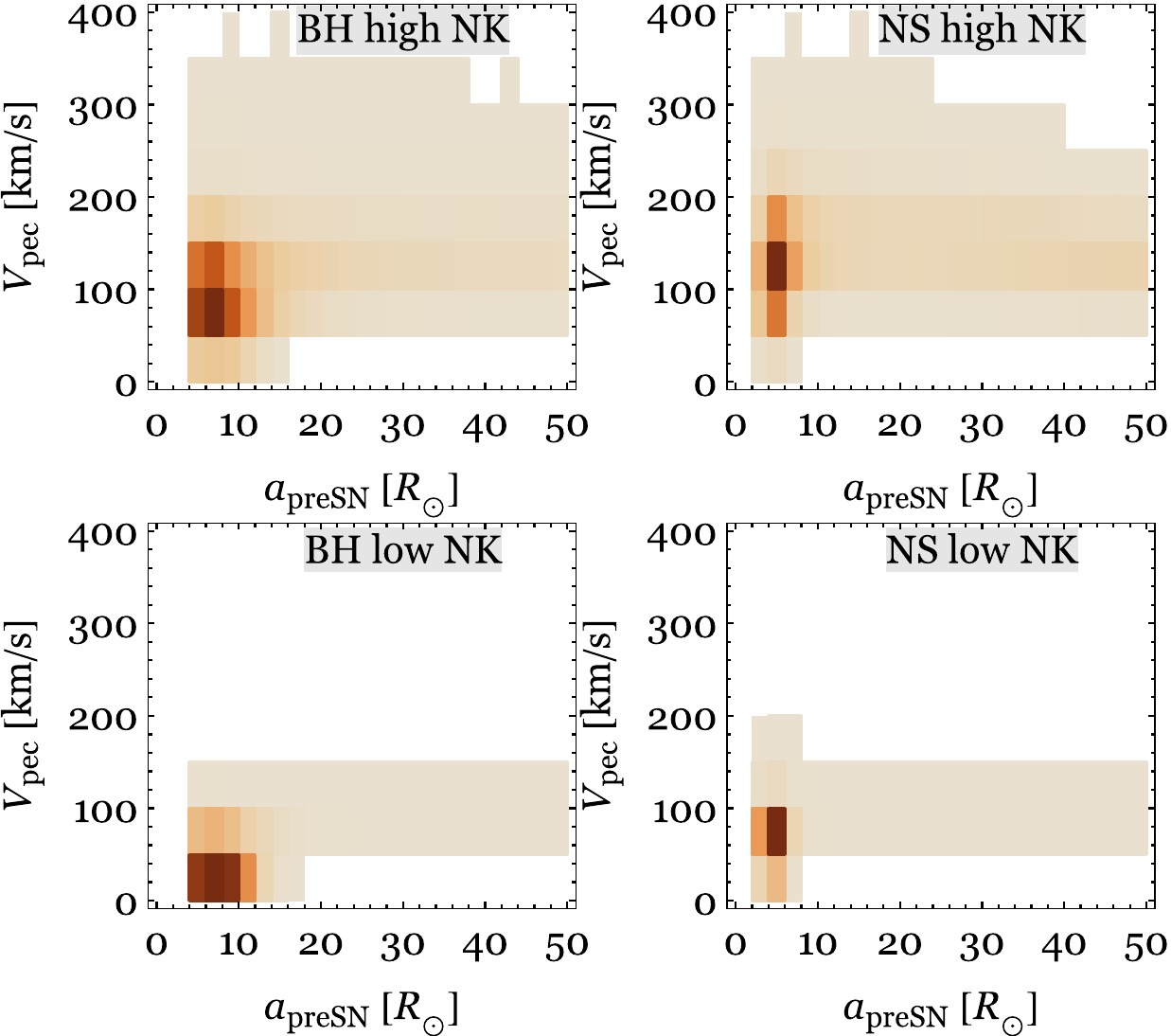}
\includegraphics[width=0.5\columnwidth]{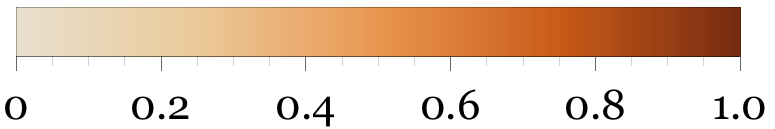}
\caption{Density plots which result from our population synthesis models showing the allowed parameter space for the peculiar velocity at birth $V_\mr{pec}$ and the orbital separation $a_\mr{preSN}$ of BH- and NS-XRBs prior to the formation of the compact object. Each panel corresponds to different assumptions on the NK. The fraction of systems in each 2-dimensional bin is shown;
darker colours correspond to a larger fraction of systems.}
\label{fig:preSN}
\end{figure}

In the models we used in Section \ref{sec:BPS},
the orbit of the binary progenitors of BH- and NS-XRBs
was chosen to be uniformly distributed in the range $[a_\mr{min}, 50]~R_\odot$.
It could be that this choice biases our results towards certain values for $V_\mr{pec}$.
To test this, we check how the distribution of the initial orbital separation of the binaries
(i.e. prior to the formation of the compact object)
varies with the magnitude of the NK and of $V_\mr{pec}$.
From Figure \ref{fig:preSN},
it is clear that the majority of the initial orbital separations are constrained to lie within a small range 
($a_\mr{preSN}\lesssim 10~R_\odot$) both for NS and BH systems,
and both for high and low NKs. Furthermore, there is no clear trend of $V_\mr{pec}$ with respect to 
$a_\mr{preSN}$.
We hence conclude that it is unlikely that the peculiar velocities $V_\mr{pec}$ would be very
much influenced if the pre-SN orbits had a distribution different from the uniform one we use in our study, 
or if they were 
drawn from a smaller range.

\subsection{Comparison with observations: BH-XRBs}
\label{s:rules_for_cumul}
\begin{figure}
\centering
\includegraphics[width=0.9\columnwidth]{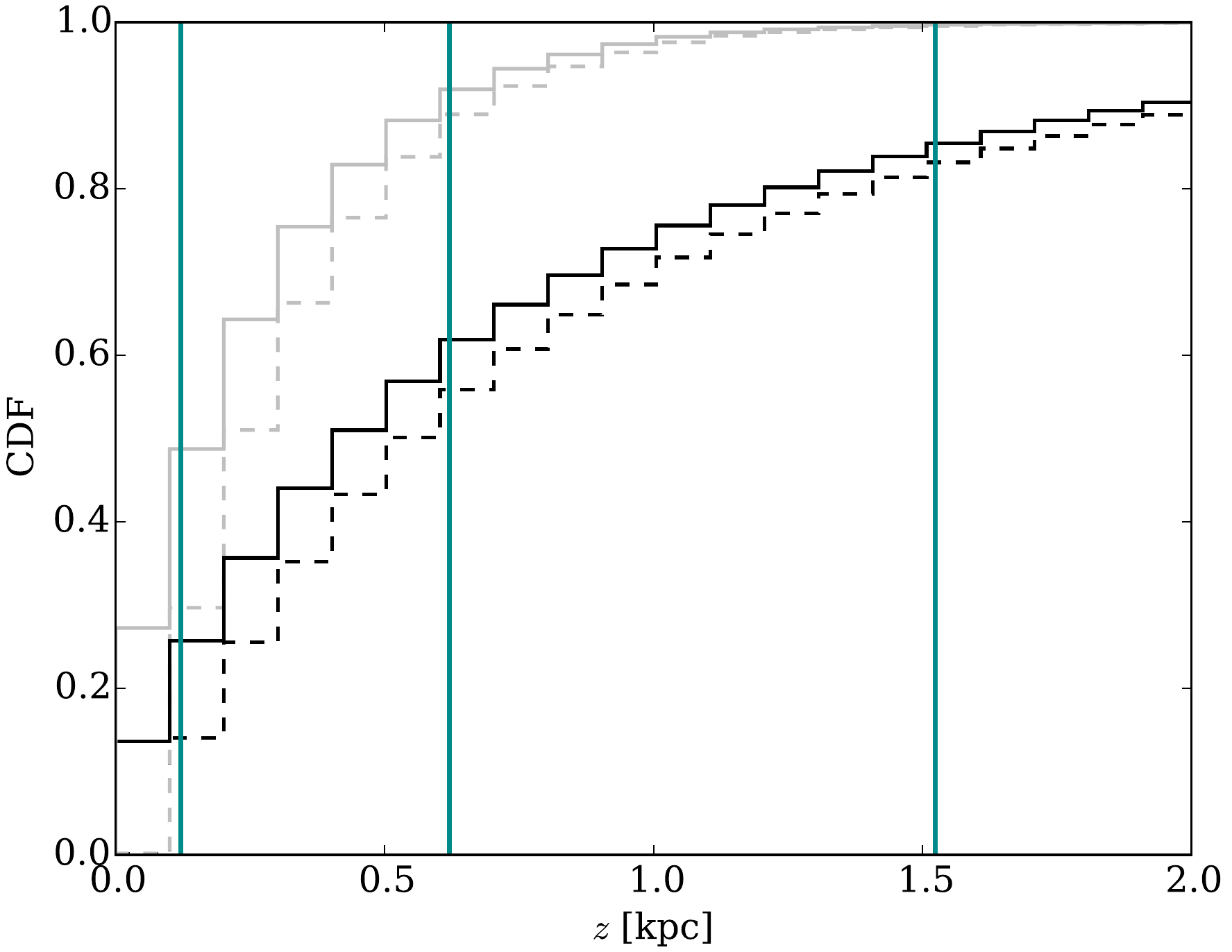}
\caption{The cumulative distribution for $|z|$ for Model 1 (black lines) and Model 2 (grey lines), in the bin $R=[8,9]$ kpc. Solid lines correspond to the whole sample of simulated binaries;
dashed lines correspond to the remaining part of the sample after the exclusion of systems close to the Galactic plane, i.e. $z\leq 0.1$ kpc. The blue vertical lines represent the observed $|z|$ of 3 BH-XRBs (from left to right: 1A 0620-00;
GRS 1009-45; XTE J1118+480).}
\label{fig:intersection}
\end{figure}

\begin{figure}
\centering
\includegraphics[width=0.9\columnwidth]{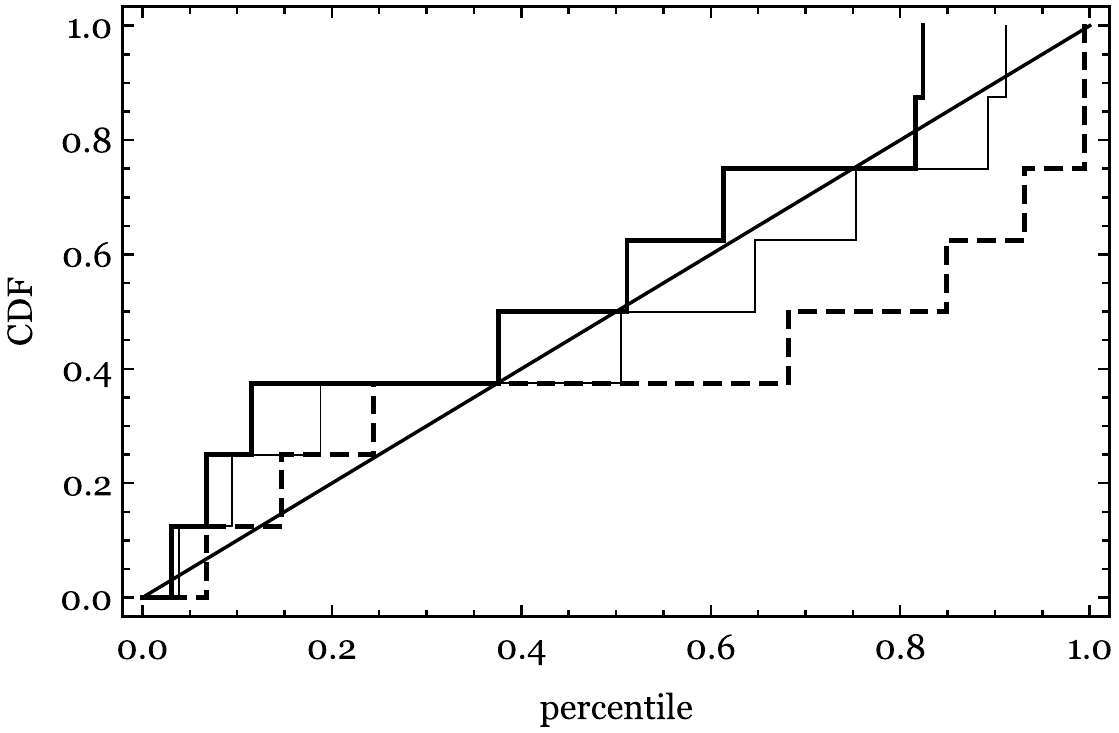}
\caption{ Cumulative distribution of the percentiles associated with short- period dynamically-confirmed BH-XRBs in Model 1 (solid line) and Model 2 (dashed line). The thin solid line is a blend of Model 1 and 2 ($50 - 50\%$). The model which fits best the observed data is the one closer to the diagonal line.}
\label{fig:obs1}
\end{figure}

\begin{figure}
\centering
\includegraphics[width=0.9\columnwidth]{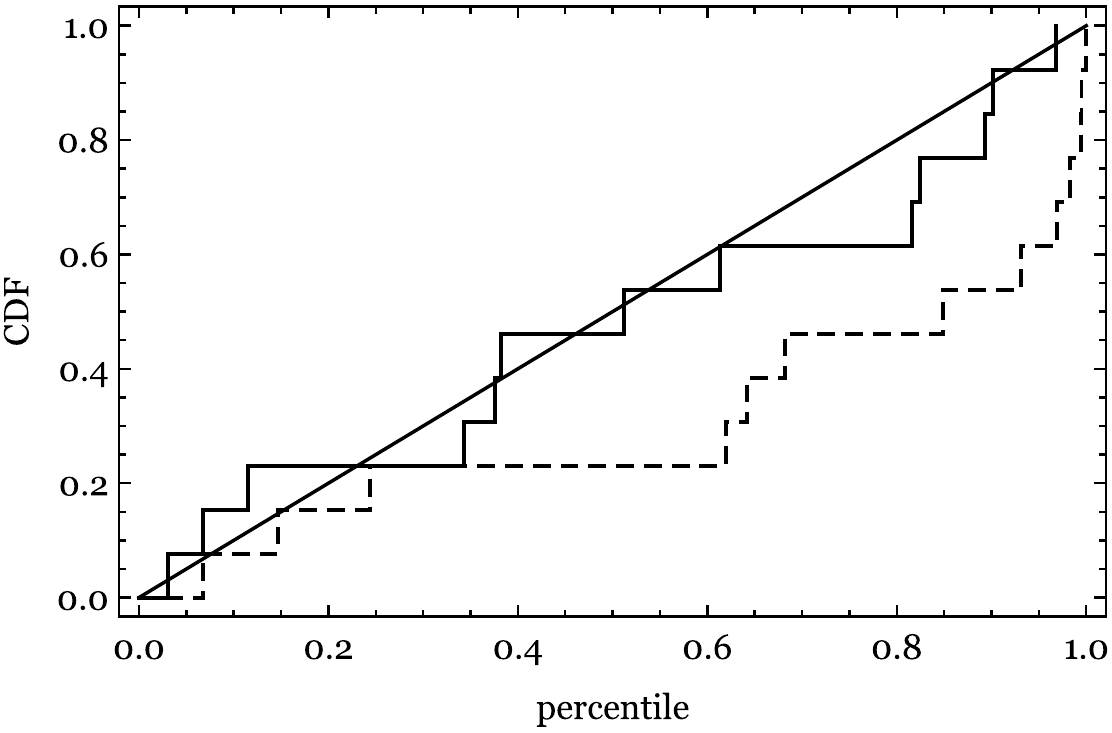}
\caption{Cumulative distribution of the percentiles associated with short- period dynamically-confirmed and candidate BH-XRBs in Model 1 (solid line) and Model 2 (dashed line). The model which fits best the observed data is the one closer to the diagonal line.}
\label{fig:obs2}
\end{figure}

\begin{figure}
\centering
\includegraphics[width=0.9\columnwidth]{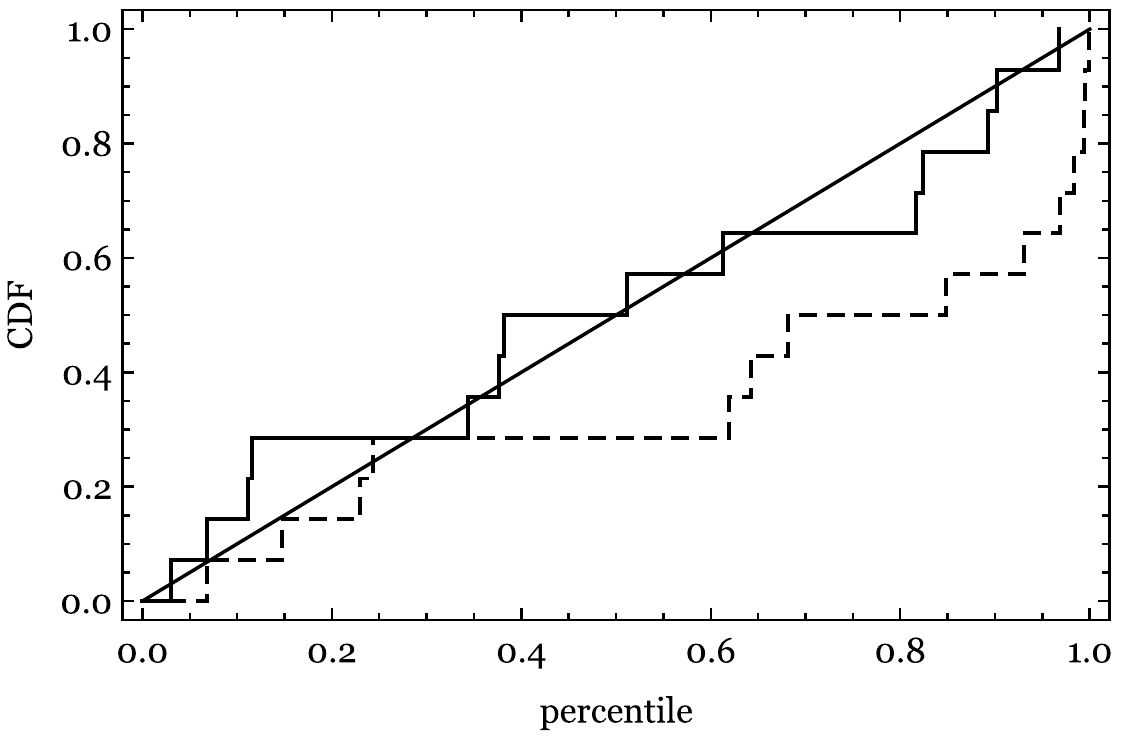}
\caption{Cumulative distribution of the percentiles associated with the whole sample of BH-XRBs in Model 1 (solid line) and Model 2 (dashed line). The model which fits best the observed data is the one closer to the diagonal line.}
\label{fig:obs4}
\end{figure}

We now turn to the comparison of the different models
with the observed BH-XRBs.
In order to
compare the simulations with the observed systems, we note that every subgroup of BH binaries of Table \ref{tab:systems}
gives rise to a certain 2D distribution in $R$ and $z$.
One way of proceeding would be to compare the 2D simulated distribution with the 2D observed one.
We compare the data with the simulated populations dividing the Galaxy into $1$ kpc-wide bins along the $R$-direction.
This allows to 
account for the fact that the Galactic potential is a strong function of the position
in the disk,
as we showed in Section \ref{sec:Effect}.
For every $R$-bin,
we compute the cumulative distribution function (CDF) of the height $z$ above the Galactic plane 
based on the population synthesis results within Model 1 and Model 2
(see as an example black and grey lines in Figure \ref{fig:intersection}, for the bin: $R=[8,9]$ kpc).
Then we calculate where in the cumulative distribution the observed systems lie
(see as an example the intersection between the blue vertical lines and the CDFs in Figure \ref{fig:intersection}). 
In such a way, we obtain a list of percentiles.
If the model is correct,
we expect these percentiles to be drawn from the uniform distribution.
We note that we have removed from our comparison those sources located in the bulge of the Galaxy
(i.e. H 1705-250 and MAXI J1659-152),
which could have had a different origin rather than having formed in the plane
(see Section \ref{sec:Effect}).
We plot the cumulative distribution of these percentiles in  
Figures
\ref{fig:obs1} (short-period confirmed BH-XRBs), \ref{fig:obs2} (short-period confirmed + candidates), and \ref{fig:obs4} (whole sample).
In the figures,
the solid lines correspond to Model 1 and the dashed lines correspond to Model 2.
We also consider a model  which consists of a superposition of Model 1 and Model 2
in equal parts (see thin solid in Figure \ref{fig:obs1}, in the case of the short-period confirmed BH-XRBs). The model which fits best is the one which comes closer to the diagonal line (that
represents the cumulative of a uniform distribution).
In all three cases,
a high NK distribution is the most preferable one.

We perform 
a Kolmogorov-Smirnov (KS) test to measure how close
is the distribution of percentiles to the diagonal line of Figures \ref{fig:obs1}, \ref{fig:obs2}, \ref{fig:obs4}. We summarise the $D$-values and their corresponding probabilities 
in Table \ref{tab:Dvalues} for every subgroups of BH-XRBs.
For each of the sub-groups the high-NK model fits the
  data best, although in the group with confirmed BHs only,
  the low-NK is also consistent with the data. 
For the confirmed+candidate short-period systems as well as for the whole sample,
 the low-NK model is
  inconsistent. 
  Interestingly, the model in which the BHs receive both
  low and high NKs, fits the data best for the confirmed systems.

\begin{figure}
\centering
\includegraphics[width=0.9\columnwidth]{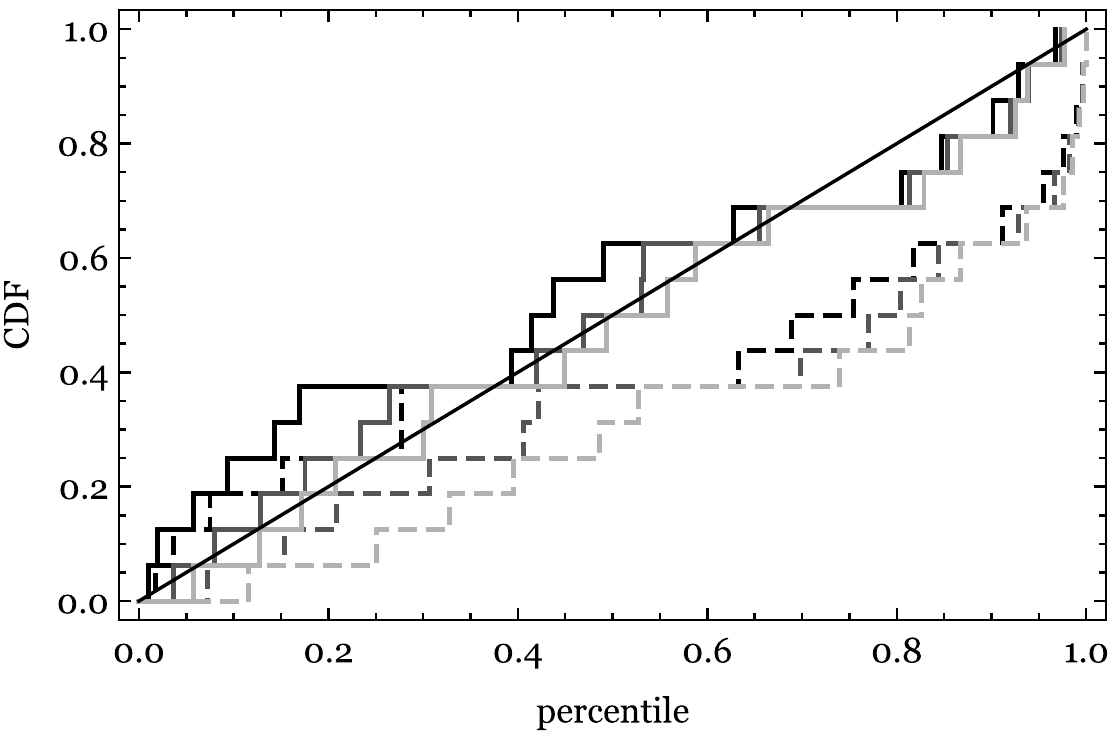}
\caption{Cumulative distribution of the percentiles associated with the whole sample of BH-XRBs in Model 1 (solid lines) and Model 2 (dashed lines) when assuming a different fraction of systems in the Galactic plane: $f_\mr{disc} = 0.1$ (black lines),
$0.5$ (darker grey lines), or $0.9$ (lighter grey lines). The model which fits best the observed data is the one closer to the diagonal line.}
\label{fig:obs5}
\end{figure}

{{In these results, we have excluded all the systems in the plane (both observed and simulated).
An accurate modelling of the obscured systems would require a model for the Galactic extinction in and out of the plane
combined with a model for the optical/NIR magnitudes of BH-XRBs in their quiescent state.
As a first step, we simplistically model the observational effects near the Galactic plane including a certain fraction of those simulated points which end up in the Galactic disc (at $z\leq 1 $ kpc): either $f_\mr{disc}=0.1$, or $0.5$, or $0.9$. We compare the Galactic distribution of these simulated binaries with the distribution of the whole sample of binaries,
 including this time the obscured sources GRS 1915+105 and V404 Cyg as well. The results are presented in Figure \ref{fig:obs5} and Table \ref{tab:Dvalues}. Also when including the obscured systems, the high-kick model is the most successful in reproducing the observed binaries.}}

\begin{table*}
\caption{D-values of the KS-test for different systems and in the different models: Model 1 (i.e. high NK), Model 2 (i.e. low NK), and a model made
of a superposition of the high- and low-NK in equal parts.}
\label{tab:Dvalues}
\begin{tabular}{l c c c c c}
\hline
Subgroup & High NK & Low NK    & 50-50    & N  & Fig. \\
               & D (P)   & D (P)     & D(P)    &    & \\
\hline
BH-XRBs, short P., confirmed & $0.26$ $(0.57)$ & $0.34$ $(0.24)$ &$0.19$ $(0.92)$ & 8  & \ref{fig:obs1} \\
BH-XRBs, short P., confirmed+candidates & $0.20$  $(0.61)$ & $0.39$ $(0.03)$ & $0.28$ $(0.22)$ & 13 & \ref{fig:obs2} \\
BH-XRBs, whole sample &  $0.17$ $(0.77)$ &$0.36$ $(0.04)$ & $0.26$ $(0.24)$ & 14 & \ref{fig:obs4} \\
BH-XRBs, whole sample, $f_\mr{disc}=0.1$ &  $0.20$ $(0.46)$ &$0.29$ $(0.12)$ & $0.19$ $(0.54)$ & 16 & \ref{fig:obs5} \\
BH-XRBs, whole sample, $f_\mr{disc}=0.5$ &  $0.13$ $(0.96)$ &$0.33$ $(0.04)$ & $0.20$ $(0.47)$ & 16 & \ref{fig:obs5} \\

BH-XRBs, whole sample, $f_\mr{disc}=0.9$ &  $0.14$ $(0.91)$ &$0.37$ $(0.01)$ & $0.22$ $(0.37)$ & 16 & \ref{fig:obs5} \\
\hline
NS-XRBs                &  $0.39$ $(0.06)$ & $0.63$ $(0.00)$ & - &  10 &  \ref{fig:obsNS} \\
\hline
\end{tabular}
\newline
\end{table*}

\subsubsection{Effect of the distance uncertainty}
The distance $d$ to a BH-XRB is typically estimated by measuring the 
apparent magnitude of the companion star in a certain colour band,
and computing its absolute magnitude.
Once an estimate of the reddening towards the source is known and the
spectral type of the donor star is clearly identified,
the distance can be calculated.
In the best case scenario, one
would have the apparent magnitude of the source in different bands,
and then would compute the scatter between the derived distances
as estimate of the distance uncertainty. We expect such 
uncertainties  to follow a Gaussian distribution.
However,
in case a range of spectral types is equally probable,
we expect the errors on the distance to be distributed more uniformly.
To investigate the influence
of the uncertainty in the distance,
and since for most of the literature there is no easy way of determining the type of error distribution,
we randomly generate 100 values for the distance to each BH-XRB,
either distributed as a Gaussian (with $\sigma$ equal to the distance uncertainty $\delta$) or as a uniform distribution in the range $(d-\delta, d+\delta)$.
Such errors can cause a binary to move from one $R$-bin to the adjacent one, affecting the percentile values.
However, 
we find that there is no systematic shift that would make low NKs fit best the observed data,
$\delta$ being smaller than the discrepancy between the two distributions.

\subsection{Comparison with observations: NS-XRBs}
\label{sec:ObsNS}
\begin{figure}
\centering
\includegraphics[width=0.9\columnwidth]{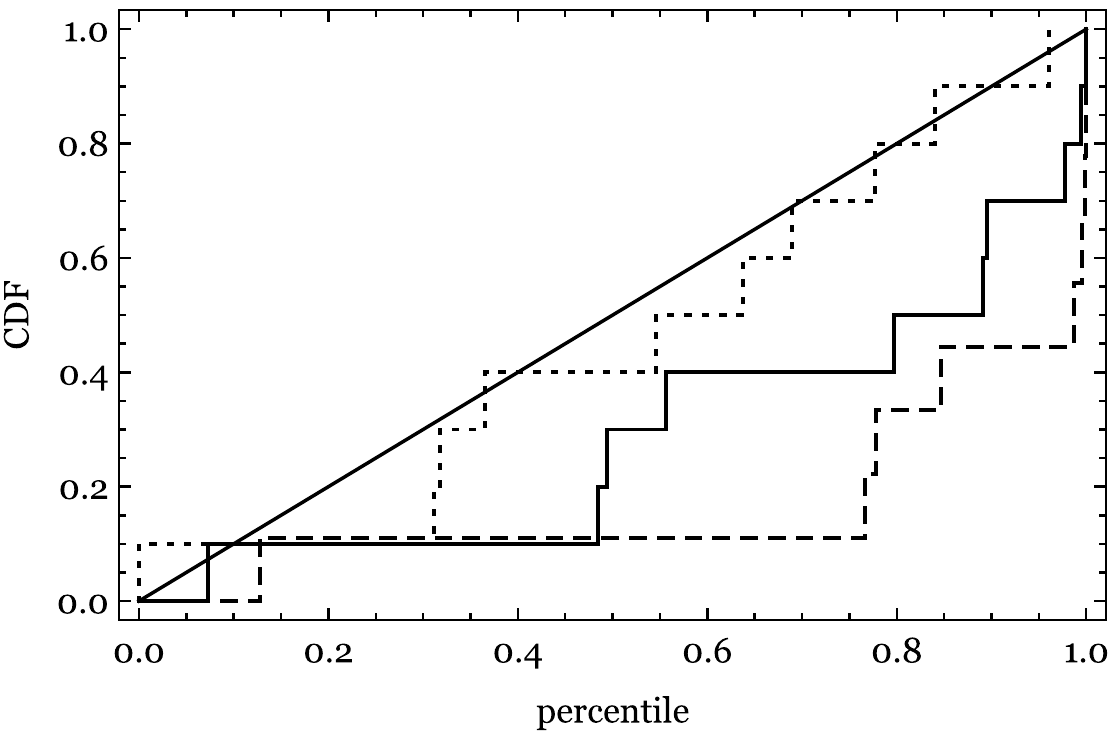}
\caption{Cumulative distribution of the percentiles associated with short- period NS-XRBs in Model 5 (solid line) and Model 6 (dashed line). Dotted line is when the NK is drawn from the Hobbs distribution. The model which fits best the observed data is the one closer to the diagonal line.}
\label{fig:obsNS}
\end{figure}

We compare the observed $z$ distribution of NS systems with the distribution 
of the two simulated population of NS-XRBs in the context of
Model 5 and Model 6.
We perform the comparison in the same way we did for BH-systems in Section \ref{s:rules_for_cumul}.
From Figure \ref{fig:obsNS} we see that none of the distributions (solid and dashed lines)
fits the data.
{{Our goal is not to calibrate the NS NK distribution from the NS-XRB
population, nor from a population model of radio pulsars (cf. \citealt{1997A&A...322..477H}). Nevertheless, we can note that the observed population of NS-XRBs seems to be consistent with NKs larger than $\approx 100$ km/s. This is in line with the catalogue of pulsar proper motions by \citealt{2005MNRAS.360..974H}, who inferred a mean pulsar birth velocity of $\approx 400$ km/s. 
However, the derivation of pulsar velocities from the measured proper motions has to be taken with caution, because of the possible uncertainties in the proper motion measurements as well as in the distance measurements. 
More in general, underestimating proper motion measurement errors can lead to an overestimate of pulsar velocities, as noted by \citet{1997A&A...322..127H}. The distance to a pulsar is typically estimated through parallax. \citet{2016A&A...591A.123I} showed that a more proper Bayesian approach to calculate the distance probability function from a single parallax measurement has to be used. Such method has not been applied yet to the whole population of pulsars. }}

We show the results of the KS-test for NS systems in Table \ref{tab:Dvalues}: both models have large $D$-values.

{{For an illustrative purpose, we also compare the observed population of NS-XRBs to a simulated one
in which the NK is drawn from a Maxwellian distribution with $\sigma=265$ km/s (\citealt{2005MNRAS.360..974H}).
The results of the KS test favours this distribution: $(D, p)_\mr{Hobbs}=(0.21, 0.72)$; see dotted line in Figure \ref{fig:obsNS}}}.

{{We note that we did not include the long-period NS-XRBs to our study as in the sample of NS-XRBs from \citealt{2004MNRAS.354..355J} that we are using, there is only one long-period system with a low-mass companion, Cygnus X-2.}}

\section{Discussion} 
\label{sec:Discussion}
\begin{enumerate}
\item In our models of Section \ref{sec:BPS},
we have assumed an ejected mass at BH formation
of $0$ or $4~M_\odot$. Taking $M_\mr{ej}=8~M_\odot$ would not greatly affect the scale height of the binaries
in the case of a high NK distribution,
since the typical peculiar velocities are comparable to the case when $4~M_\odot$ are ejected
(see dotted line in Figure \ref{fig:vsys}). This is due to the fact that a high ejected-mass is compatible
only with the lower-velocity tail of the NK distribution in order for the binary to stay bound.
In the case of a low NK distribution,
the higher ejected-mass has a greater effect on the average peculiar velocity (see dashed-dotted line in Figure \ref{fig:vsys}). 
However, the NS systems velocities are still larger.
 In order for BH systems to have the same peculiar velocity as NS systems,
we would need that $V_\mr{MLK, BH} = V_\mr{MLK, NS}+ (1/2) \times V_\mr{NK}$, which follows from the expression for $V_\mr{pec}$  (equation \ref{eq:PEC}), imposing that $V_\mr{pec,BH}=V_\mr{pec,NS}$
and assuming that $V_\mr{NK,x}=V_\mr{NK}$.
This would constraint $M_\mr{ej}$ to be much larger than what is allowed for the binary to stay bound.
\item {{In the modelling of the progenitors of BH- and NS-XRBs, we have assumed a flat distribution of the initial orbital separation in the range $[a_\mr{min}, 50]~R_\odot$ (Section \ref{sec:BPS}). With this choice, we are including all possible pre-SN orbital separations. In~\citet{2015MNRAS.453.3341R} we verified that larger separations
do not contribute to the final separation of the binary (see their Figure 13).
This is due to the fact that the strength of the coupling between tides and magnetic braking, which is responsible for the shrinking 
of a binary to short orbital periods, decreases strongly with larger orbital separations. For long-period BH-XRBs,
this choice is also acceptable, as none of the observed binaries in our sample have an orbital separation larger than $50~R_\odot$.
There could be of course cases in which the orbit in the post-SN phase is highly eccentric and very wide, but these cases would contribute only to a minority of the systems.
More importantly, we have found that the NK distribution does not in
fact depend on the pre-SN orbital period (see Figure \ref{fig:preSN}, where there is no trend of the orbital separation depending on the NK). We can also compare the circularised orbital period distribution after the SN in our models,
$P_\mr{orb,circ}$, 
with the one in \citealt{2003ApJ...597.1036P}, who did detailed evolutionary calculations of NS-LMXBs. In our models, $P_\mr{orb,circ}$
ranges from $\approx 0.15$ days to $\approx 12$ days, which is compatible with the range shown by \citealt{2003ApJ...597.1036P}
in their Figure 1.}}
\item {{We have assumed that the companion of BHs and NSs are stars with an initial mass of $1~M_\odot$. \citet{2003ApJ...597.1036P}
argued that the majority of LMXBs have likely originated from binaries with an intermediate mass companion ($\approx 2-3~M_\odot$). We have checked how our results would be affected when taking a companion of initial mass: $3~M_\odot$. The peculiar velocity right after the BH formation (see Figure \ref{fig:vsys}) would decrease due to the larger binary mass: by a factor of $\approx 0.6$ on average. This implies that the NK would need to be even larger in order for the simulated systems to match the observed ones.}}
\item The two Maxwellian distributions used in Section \ref{sec:BPS} do not correspond to the real physical distributions,
but rather are representative of two complementary distributions,
one generating large kicks, the other generating low kicks.
The choice of two distributions peaked at two different velocities
serves the purpose of analysing how close in magnitude are the velocities received by BHs with 
respect to the velocities received by NSs.
{{The NK distribution of NSs 
can be estimated via proper motion studies of pulsars.
The pulsar birth speed distribution has been estimated
as a Maxwellian distribution with $\sigma = 265$ km/s by \citealt{2005MNRAS.360..974H}; however, one should bear in mind the caveats discussed in Section \ref{sec:ObsNS}.}}
For BHs, the number of sources with measured 3D space velocity ($5$; see \citealt{2014PASA...31...16M}) is not sufficient 
to allow for a calibration of their NK distribution.
\item When comparing the observed BH-XRBs with the synthetic BH-XRBs which result from our population synthesis,
we found that the population in which BHs receive high NKs best fit the observed data.
This conclusion gains strength when including in the comparison between observations and simulated population the sources located in the bulge of the Galaxy (i.e. H 1705-250 and MAXI J1659-152).
In this case, the KS-values for the short-period confirmed BH-XRB sample are:
$(D, p)_\mr{highNK}=(0.22, 0.74)$, $(D, p)_\mr{lowNK}=(0.40, 0.08)$,
and for the short-period confirmed + candidates BH-XRB sample are:
$(D, p)_\mr{highNK}=(0.28, 0.15)$, $(D, p)_\mr{lowNK}=(0.46, 0.00)$.
\item {{We did not include the long-period NS-XRBs in our study. We can still investigate how much long- and short-period NS-XRBs differ when it comes to the peculiar velocity after the SN, and hence how they differ in terms of the scale height above the Galactic plane. We build a population of binaries which evolve into long-period NS-XRBs along the lines of the computational method used for long-period BH-XRBs (see Section \ref{sec:BPS}). We find that the long-period systems have slightly lower peculiar velocities, which results in a slightly lower scale height (by a factor of $\approx 0.8$, for every radial distance). This is due to the fact that the binary, having lower binding energy, can only survive lower kicks.}}
\item {{As it was mentioned in the introduction, there is evidence for some NSs receiving low kicks at birth: NSs residing in double-NS systems (\citealt{2010ApJ...721.1689W}; \citealt{2016MNRAS.456.4089B}; \citealt{2016arXiv161105366C}) and NSs hosted in a subset of high-mass X-ray binaries (HMXBs; \citealt{2002ApJ...574..364P}). It was suggested that electron-capture SNe are too fast for large asymmetries to develop, resulting in a modest kick, of a few $10$ km/s at maximum. Such kicks are of the order of the low natal kick model we took in our binary population synthesis of Section \ref{sec:BPS} through which we found (Section \ref{sec:ObsNS}) that high-kick models best fit the Galactic distribution of NS-LMXBs: i.e. we did not find any evidence for NSs born in an electron-capture SN. This might be due to a different evolutionary path for the progenitors of NS-LMXBs. Several studies have investigated the type of SN event (either a standard collapse of the iron core or an electron capture SN) as a function of the evolutionary state of the helium star progenitor of the NS as well as the characteristics of the binary orbit. \citealt{2015MNRAS.451.2123T} performed detailed evolutionary sequences of binaries hosting an NS and a helium star. They found that helium stars
with an initial mass of $2.6-2.95~M_\odot$
could be stripped of a significant fraction of their mass through case BB mass transfer  towards an NS companion when the helium star expands as a giant. The low-mass core would then form an NS via electron-capture SN, 
provided the initial orbital period is sufficiently wide. Shorter orbital period would instead result in a white dwarf (WD). We can extrapolate this finding 
to the short-period systems formed by an NS and a low-mass companion studied in our work. However, we must bear in mind that 
the stripping effect is less understood for a giant helium star experiencing mass transfer to a low-mass main-sequence star, as indeed mentioned by \citealt{2015MNRAS.451.2123T}.
Thus we conclude that, heretofore, there is no strong theoretical support for a preference for electron-capture SN in the progenitors of NS-LMXBs. We also wish to note that the study by \citealt{1998ApJ...493..351K} highlighted the impossibility of forming short-period (less than one day) NS-LMXBs without large natal kicks at birth (where for large they chose a Maxwellian distribution with an average kick of 300 km/s).}}
\end{enumerate}

\subsection{A note on our KS-test}
In assessing the quality of the fit of our simulations, we used the
classical application of the KS-test. In order to test its validity and
determine the {\emph{power}} of the test in distinguishing the two
hypothesis we draw samples of various sizes from the simulated
populations, and we calculate the $D$-value distribution of each of
these samples, according to the rules described in
Section~\ref{s:rules_for_cumul}. We find that the probabilities follow
the classical KS test (as expected) to an accuracy ($5\%$) that is
comparable to the Poisson noise in our simulations ($\approx 3\%$). 
More interesting is the measurement of how often we obtain $D$-values
smaller then the ones we measured for our observed samples when
\emph{testing the wrong hypothesis} - i.e. when using a sample drawn
from the high (low) NK synthetic population and testing the low (high)
NK hypothesis (also known as false negative rate). For the BH case, we
find $D$-values smaller than the ones in Table \ref{tab:Dvalues} in
less than $\beta \approx 10\%$ of the cases, for the high NK hypothesis,
  and in more than $\beta \approx 30\%$ for the low NK hypothesis.
 A particularly interesting fact is that $\beta = 0.015$ for
short-period confirmed + candidate BH-XRBs. If we accept the $\alpha =
3\%$ confidence level and use the standard 4-to-1 weighting (i.e.
$4\alpha = \beta$ ), the test we developed has enough
power to distinguish between the high- and low-NK hypothesis in this
case, and that the high-NK hypothesis is clearly preferable. This is a
{\emph{non}}-expected result given the small number of object, and it
can potentially dissolve if some of the BH candidates turn out to be
NS-XRBs. To calculate what is the optimal number of observed systems to
decrease such rate $\beta$, we draw samples of various sizes from the
population synthesis results of Model 1 and we test the low-NK
hypothesis, and vice-versa.  To decrease this rate to the level that in
$95\%$ of cases we obtain $\beta < 1\%$, we find that it is necessary
to increase the size of the observed sample to $\approx 40$ systems,
both in the BH and NS case.

\section{Conclusions}
\label{sec:conclusion}
In this work we performed a binary population synthesis study of BH- and NS-XRBs,
tracing their binary evolution from the moment of compact object formation until the observed phase of mass transfer,
and integrated their orbits in the Galaxy. The main goal was to investigate whether different assumptions
on compact object formation manifest themselves in the Galactic distribution of the binaries.
We found that these assumptions do affect the scale height of the binaries, which we quantified through their $z_\mr{rms}$.
In particular, we found that if BHs and NSs receive the same NK at birth,
NSs would still have a larger scale height above the Galactic plane,
due to the fact that their systemic velocities acquired when the compact object is formed are typically larger,
their total binary mass being smaller. The larger scale height of NS-XRBs with respect to BH-XRBs
is clearly seen also in the observed populations.
We also found a clear trend for both populations of increasing scale height for larger Galactocentric radii,
which should manifest itself,
but which is not clearly observed in the current populations
(see Figure \ref{fig:Space}).

The main outcome of this study is that when analysing the $z$-distribution of the observed BH systems
as a function of $R$, the simulated population in which at least some BHs receive a (relatively) high NK ($\sim 100$ km/s) fits the data best.
This is in agreement with previous findings by \citet{2012MNRAS.425.2799R},
who
compared the observed and simulated populations of BH-XRBs only in the $z$-direction,
whereas we compare the 2D distributions, accounting for how the binaries are distributed along the $R$-direction as well.
Furthermore,
we increased the sample of sources adding $6$ BH candidates, updated their distances according to the recently published BH catalogue of \citet{2016A&A...587A..61C},
and followed the binary evolution of the binaries in a detailed way (accounting in particular for magnetic braking and tides).

In this work we also checked numerically the validity of a simple one-dimensional analytical estimate for the peculiar velocity at birth of BH-XRBs
which we used in our previous works \citet{2012MNRAS.425.2799R}
and \citet{2015MNRAS.453.3341R}. We found that this estimate is less reliable for some gravitational potentials for sources
in the bulge of the Galaxy, i.e at $R\leq1$ kpc.
This was also shown by \citet{2016MNRAS.456..578M},
who studied the kinematics of H 1705-250, a BH-XRB close to the Galactic bulge.
However, the estimate is robust for systems at Galactocentric radii larger than $1$ kpc.
\citet{2015MNRAS.453.3341R} 
followed the binary evolution of seven short-period BH-XRBs and estimated their minimal peculiar velocity at birth,
to conclude that two out of the seven sources were
consistent with a high (or relatively high) NK at birth. This conclusion remains valid even in view of the current analysis.

\citet{2004MNRAS.354..355J} found that the rms-value of the distance to the Galactic plane for BH-XRBs was similar to that of NS-XRBs. This 
was suggestive for 
BHs receiving a kick-velocity  at formation.
We revised the distances and updated the sample of BH-XRBs using the catalogue from 
\citet{2016A&A...587A..61C} and we found that NS systems have a larger scale height than BH systems,
a trait which is also present in the simulated populations.

Finally, we found that the comparison of the
  data to our simulations is limited by the small number of observed
  BH-XRBs, and thus that more systems should be found to
  determine in more detail the NK that BHs receive.
In this respect, the possible future discovery of new BH transients with Gaia (\citealt{2014SSRv..183..477M}),
and through dedicated surveys
such as the Galactic Bulge Survey (\citealt{2011ApJS..194...18J}), are promising.

\section{Acknowledgments}
SR is thankful to Jo Bovy for answering her questions concerning Galpy
and for useful comments on the manuscript.
SR wishes to warmly thank her office mates (Pim van Oirschot, Martha Irene Saladino, Laura Rossetto, Thomas Wevers) at Radboud University for their great moral support
during the hectic times of finishing up her thesis. SR wishes to acknowledge Melvyn B. Davies with whom she started her research on black hole natal kicks.
This research has
made use of NASA\textquotesingle s Astrophysics Data System.

\end{document}